\documentclass[a4paper]{llncs}
\pdfoutput=1
\usepackage[a4paper]{geometry}
\usepackage{commath}
\usepackage{amssymb}
\usepackage{url,lineno}
\usepackage{natbib}
\usepackage{graphicx}

\urldef{\mailsa}\path|L.Berthouze@sussex.ac.uk|

\newcommand{\keywords}[1]{\par\addvspace\baselineskip\noindent\keywordname\enspace\ignorespaces#1}

\begin{document}

\title{Markers of criticality in phase synchronisation}
\author{Maria Botcharova$^{1,2}$, Simon F. Farmer$^{2,3}$ \& Luc Berthouze$^{4,5,\ast}$}
\institute{Centre for Mathematics and Physics in the Life Sciences and Experimental Biology, University College London, UK \and
Institute of Neurology, University College London, UK \and
The National Hospital for Neurology and Neurosurgery, London, UK \and
Centre for Computational Neuroscience and Robotics, University of Sussex, Falmer, UK \and
Institute of Child Health, London, University College London, UK \\
$\ast$ Corresponding author: \mailsa}

\maketitle

\begin{abstract}
The concept of the brain as a critical dynamical system is very attractive because systems close to criticality are thought to maximise their dynamic range of information processing and communication. To date, there have been two key experimental observations in support of this hypothesis: i) neuronal avalanches with power law distribution of size and ii) long-range temporal correlations (LRTCs) in the amplitude of neural oscillations. The case for how these maximise dynamic range of information processing and communication is still being made and because a significant substrate for information coding and transmission is neural synchrony it is of interest to link synchronization measures with those of criticality. 
We propose a framework for characterising criticality in synchronisation based on a new metric of phase synchronisation (rate of change of phase difference) and a set of methods we have developed centred around the detection of LRTCs. We test this framework against two classical models of criticality (Ising and Kuramoto) and recently described variants of these models aimed to more closely represent human brain dynamics. From these simulations we determine the parameters at which these systems show evidence of LRTCs in phase synchronisation. We demonstrate proof of principle by analysing pairs of human simultaneous EEG and EMG time series, suggesting that LRTCs of corticomuscular phase synchronisation can be detected in the resting state and experimentally manipulated. The existence of LRTCs in fluctuations of phase synchronisation suggests that these fluctuations are governed by non-local behaviour, with all scales contributing to system behaviour. This has important implications regarding the conditions under which one should expect to see LRTCs in phase synchronisation. Specifically, brain resting states may exhibit LRTCs reflecting a state of readiness facilitating rapid task-dependent shifts towards and away from synchronous states that abolish LRTCs.
\end{abstract}

\section{Introduction}
The concept of the brain as a dynamical system close to a critical regime is attractive because systems close to criticality are thought to maximise their dynamic range of information processing and communication, show efficiency in transmitting information and a readiness to respond to change~\citep{chialvo,shew,stam, beggsandplenz,shew13,linkenkaer01,sornette,timme,werner2009,linkenkaer04,meisel,kinouchi}. 

A number of modeling studies have shed important light on the behaviour of neurally inspired systems close to their critical dynamical range~\citep{bullmore,poil2012,shew,breakspear,daffertschofer}.   To date there have been two significant experimental observations suggesting that the brain may operate at, or near, criticality.  These are: i) the discovery that the spatio-temporal distribution of spontaneous neural firing statistics can be characterised as neuronal avalanches with a power law distribution of avalanche size~\citep{beggsandplenz} and ii) the presence of long-range temporal correlations (LRTCs) in the amplitude fluctuations of neural oscillations, typically bandpassed MEG or EEG~\citep{linkenkaer01, hardstone}. The mechanisms by which  avalanches and LRTCs of oscillation amplitude may maximise the dynamic range of information processing and communication are still to be fully understood and experimental and computational neuroscience data linking the two phenomena are only just beginning to emerge~\citep{plenzandchialvo,poil2012} 

Population coding approaches to neuronal information storage and transmission show that both changes in the firing rate and changes in neuronal synchronisation and desynchronisation of action potentials are required to indicate changes in signal salience~\citep{baker2001,schoffelen,pfurtscheller1977,Pfurtscheller1992,singer}. At a coarser spatio-temporal scale, extracellular brain signals (local field potentials, corticography, EEG and MEG), which depend on recordings within the brain, at the brain surface and at the scalp are observed to be quasi-oscillatory (brain oscillations) and in the resting state contain spectral peaks within distinct frequency bands sitting on a $1/f$ decrease in power with increasing frequency~\citep{buzsaki}. Brain oscillations both in the resting state and during task conditions show short-range and long-range synchronisation when examined both from the phase and amplitude envelope perspectives~\citep{wang}. Primarily neuroscience has focused on the detection of synchronisation between areas either at zero phase lag, or with a fixed phase delay. This is in part a consequence of the fact that the averaging necessary to extract evidence of signal correlation requires a consistent phase relationship between the two signals for at least some period of the recording. 

Importantly, neural synchronisation is weak and it fluctuates spontaneously over time. A number of experiments have shown neural synchronisation to be consistently modulated by cognitive, perceptual and motor tasks supporting the idea that synchronisation and de-synchronisation within and across frequency bands may play an important role in communication within the nervous system~\citep{buzsaki,singer,fries,schoffelen,akam,pikovsky,doesburg,farmer98,conway95,baker}. Changing synchronisation patterns may indicate an evolution in the relationship and exchange of information~\citep{pikovsky}.  Neural synchronisation can exist between nearby and distant regions, across a range of time scales, and can be characterised using a number of techniques based on time- and frequency-domain techniques as well as mutual information~\citep{buzsaki,schoffelen,siegel,james,halliday98,brittain}. 

Neuronal synchronisation occurs when the mutual influence of neurons on each other causes them to fire  close together in time.  It is favoured by oscillatory activity. Oscillators can be tipped in and out of weak synchonization through shared noise, a phenomenon first appreciated by Huygens~\citep{pikovsky}.  Therefore weak yet variable synchrony between neuronal oscillators may easily  emerge within complex and highly interactive neural networks.  In this paper the term synchronisation will be used to encapsulate both zero and fixed phase lag synchrony but also situations in which any non-trivial phase relationship exists between signals.  Importantly, we will introduce a new methodology to demonstrate that non-fixed yet non-random phase relationships between signals are present in models of critical synchronisation and we will show that, in principle, the methodogy can be applied to neural data in order to further explore the relationship between neural synchronisation and systems operating close to a critical regime.

Recent evidence supporting the idea of criticality in the dynamics of the resting state brain activity and the appreciation that synchronisation is an important extractable property of neural spatio-temporal dynamics has led researchers to ask whether neuronal synchrony can have properties consistent with a dynamical system at criticality. These approaches identify power law distributions in neural synchronisation where synchronisation has been defined as phase consistency between two thresholded time series, e.g., see the phase lock interval (PLI) measure and the  lability of global synchronisation (GLS) measure in~\citet{bullmore}.  These findings  are of considerable interest, however, the results supporting power law behaviour of PLI have been shown by the present authors to be vulnerable to data pooling and therefore may not provide robust estimates of critical synchronisation in neural time series data~\citep{botcharova12}~\citep[see also][]{shriki}. 

As discussed above, long-range temporal correlations (LRTCs) exist in dynamical systems thought to operate close to a critical regime~\citep{linkenkaer01}. They are typically identified by the autocorrelation function of the time series decaying in the form of a power law~\citep{granger}. The detrended fluctuation analysis (DFA) technique allows a characterisation of LRTCs through an exponent similar to the Hurst exponent. DFA has been widely used in order to demonstrate the presence of LRTCs in a number of natural and human phenomena~\citep[see][for examples]{stanley, bak1996, samorodnitsky, hardstone,peng94, peng95, peng95a, karmeshu,hausdorff,wang05, robinson}.  In neurophysiology, the finding of LRTCs in amplitude fluctuations of the bandpass filtered MEG and EEG~\citep{linkenkaer01,linkenkaer04} has inspired us to develop a methodological framework that can be used to to verify the presence or absence of power law scaling of detrended fluctuations and where power law scaling is present to estimate and ascertain non-trivial DFA exponents in the moment to moment fluctuations of phase synchronisation (quantified in terms of the rate of change of phase difference time series) between pairs of neuronal oscillation time series. 

The methodology is tested as follows: i) on synthetic time series where their phase difference has known temporal properties with a known DFA exponent.  Using these simulations we demonstrate the method's ability to recover known DFA exponents in the phase difference, and we test the method's robustness to additive noise in such signals; ii) the method is tested on two classical models of criticality, Ising and Kuramoto~\citep{ising,onsager,kuramoto1975, kuramotobook}, from which time series and their pairwise phase differences can be extracted.  The output of these models is examined using our method for those parameter values that determine the sub-critical, critical and super-critical regimes. The classical Kuramoto model is tuned close to the physiological $\beta$ frequency range of MEG and EEG and examined with additive noise.  We show from this analysis that a rise in DFA exponent associated with robust power law detrended fluctuation scaling occurs close to the critical regimes of both the Ising model and the Kuramoto model with noise.

We next use our methodology to examine a system of Kuramoto oscillators, operating in a range of frequencies close to the physiological $\gamma$ frequency range of MEG and EEG that are connected through a network constructed based on empirical estimations of brain connectivity parameters with time delays, noise and non-uniform connectivity~\citep{cabral}. From these simulations, we determine the parameters at which this system shows evidence of LRTCs in the rate of change of phase differences and we relate the presence of LRTCs to the network's connectivity.

Finally, we demonstrate that in principle this methodology may be applied to neurophysiological data through analysing pairs of human EEG and EMG time series.  These preliminary results suggest that LRTCs can be detected in the phase synchronisation between oscillations in human neurophysiological recordings.

We present and discuss our methodology in detail and we offer an interpretation of its results in relation to the emerging literature on neural synchrony and criticality within neural systems.  We suggest that the existence of a valid DFA exponent in fluctuations of a phase difference measure suggests that the fluctuations are governed by non-local behaviour, with all scales contributing to system's behaviour. 

\section{Materials and Methods}
\begin{figure}
\begin{center}
\includegraphics[width=13cm]{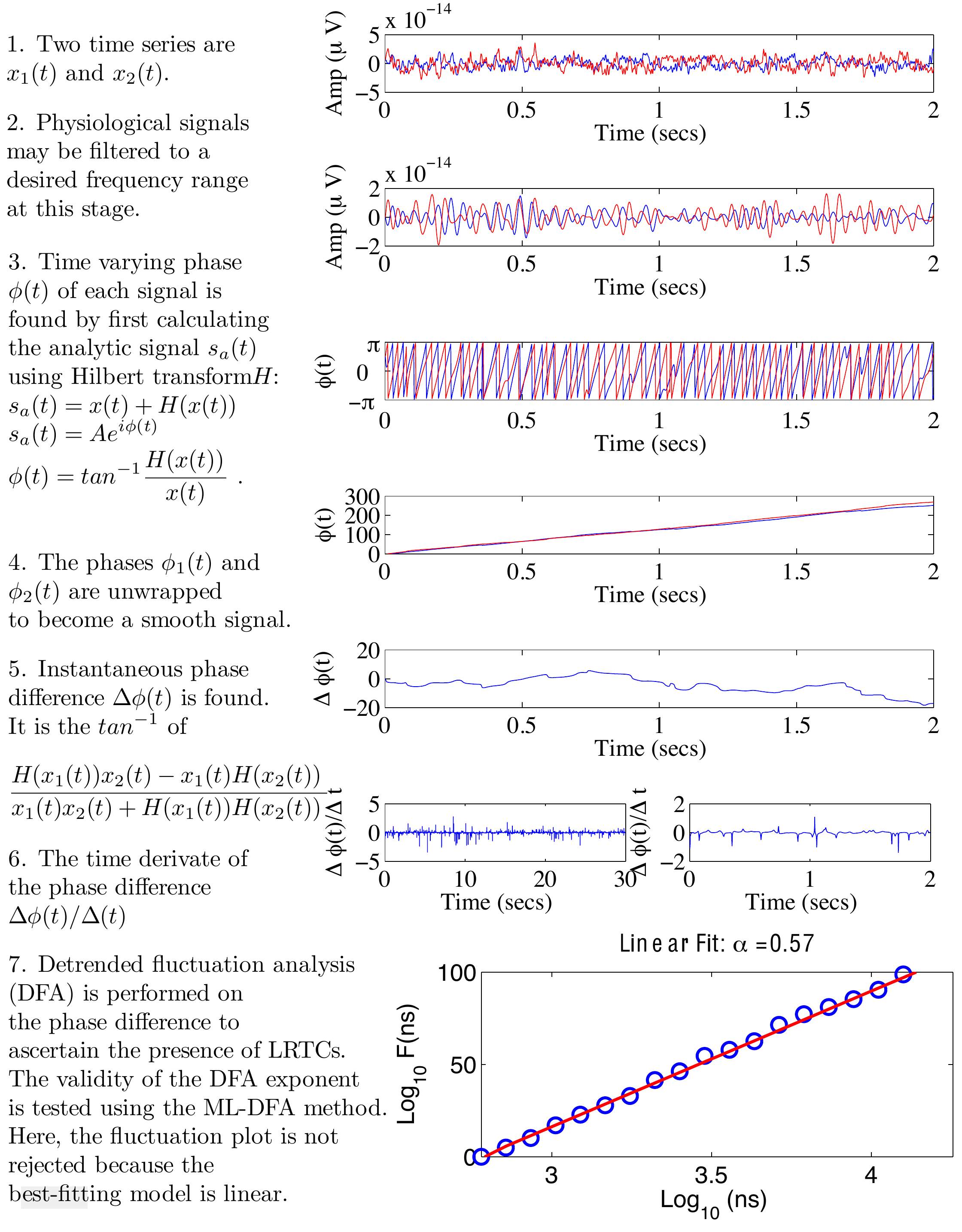}
\end{center}
 \textbf{\refstepcounter{figure}\label{fig:method} Figure~\arabic{figure}. Step-by-step illustration of the proposed method.}{ We use two sample MEG signals from the left and right motor cortex, displayed throughout panels 1-4 in red and blue, respectively. Panel 2 shows an optional bandpass filtering step. In panel 3 the instantaneous phases of the two time series are calculated using the Hilbert transform. Panel 4 shows the unwrapped phases leading to a time-varying phase difference displayed in panel 5. In panel 6, the rate of change of this phase difference is calculated. This step is illustrated using two plots, each showing a different time scale in the x-axis. These two time scales correspond to the minimum and maximum window sizes used in the DFA analysis, see Section~\ref{sec:DFA}. Panel 7 shows the resulting DFA fluctuation plot. The validity of this plot is determined using ML-DFA, see Section~\ref{sec:MLDFA}. In this case, the validity of the DFA plot was confirmed, with a DFA exponent of 0.57.}
\end{figure}

We seek to characterise the presence of long-range temporal correlations in the (time-varying) phase difference between two time series. These time series may be physiological signals such as EEG, MEG or EMG, time series extracted from a simulation or physical model, or data recorded from other natural phenomena. Below, we present the detail of the various components of our proposed methodology, including a technique used to calculate phase differences, detrended fluctuation analysis (DFA) and the recently introduced ML-DFA method for validating the output of DFA. Figure~\ref{fig:method} illustrates the application of our methodology to neurophysiological data using two sample MEG time series. We note that for these signals, we bandpassed filter the data to a frequency band of interest, however, this step will be omitted in model data considered further in the manuscript. 

\subsection{Signal Phase}\label{sec:phase}
 
The phase of a single time series $s(t)$ is calculated by first finding its analytic signal:

\begin{equation}\label{eq:analytic}
s_{a} (t) = s(t) + H \left[ s(t) \right]
\end{equation}

where $H \left[ s(t) \right]$ is the Hilbert transform:
 
\begin{equation}\label{eq:hilbert}
H[s(t)] = \mbox{p.v.} \int^{\infty}_{-\infty} s(\tau)\frac{1}{ \pi (t-\tau)} d\tau
\end{equation}

and $p.v.$ indicates that the transform is defined using the Cauchy principal value.

\subsection{Phase difference}\label{sec:phasediff}

The signal phase is defined such that it belongs to a range $\phi(t) \in [0, 2 \pi]$ or $\phi(t) \in \left[ -\pi, \pi \right] $. When a single oscillatory cycle is completed the phase returns to its starting value. A time-varying phase therefore has the properties of a sawtooth function (see panel 3 in Figure~\ref{fig:method}). In order to turn the phase into a continuous signal, the phase is unwrapped, so that at each discontinuity, a value of $2\pi$ is added to the phase~\citep{freeman04a, freeman02}. 
 
The phase difference $\phi_1(t)-\phi_2(t)$ between two different time series $s_1(t)$ and $s_2(t)$ is calculated using the respective Hilbert transform of the signals $H[s_1(t)]$ and $H[s_2(t)]$~\citep{pikovsky}:
\begin{equation}
\phi_1(t)-\phi_2(t)  = \mbox{tan}^{-1} \left\lbrace  \frac{H[s_1(t)] s_2(t) - s_1(t) H[s_2(t)]}{s_1(t)s_2(t) + H[s_1(t)] H[s_2(t)]} \right\rbrace 
\end{equation}
Full synchronisation between the two signals is indicated by a constant difference in phase over some time period~\citep{pikovsky}. The time series $\phi_1(t)-\phi_2(t)$ is an unbounded process because $\phi_1(t)$ and $\phi_2(t)$ themselves are unbounded as long as the signals $s_1(t)$ and $s_2(t)$ continue to evolve as time increases. As we shall use detrended fluctuation analysis (DFA), see Section~\ref{sec:DFA}, to assess the presence of long-range temporal correlations and DFA in its standard form assumes a bounded signal, in this paper, we characterise phase synchronisation in terms of the time derivative of the phase difference time series $\phi_1(t)-\phi_2(t)$, i.e., the rate of change of the phase difference. 

\subsection{Long-Range Temporal Correlations}

The autocorrelation function $R_{ss}(\tau)$ of a signal $s(t)$ quantifies the correlation of a signal with itself at different time lags $\tau$~\citep{priemer}, formally: 

\begin{equation}\label{eq:autocorrelation}
 R_{ss}(\tau) = \int_{\infty}^{-\infty} s(t+\tau) \bar{s}(t) dt    
\end{equation}
where $\bar{s}(t)$ is the complex conjugate of $s(t)$ and therefore $\bar{s}(t) = s(t)$ if $s(t)$ is real-valued.

In signals with short-range or no dependence~\citep{beran}, the autocorrelation function shows a rapid decay. Gaussian white noise, for example, is a signal with no temporal dependence because each successive value of the time series is independent and thus its autocorrelation function decays exponentially. In contrast, a slow decay of the autocorrelation function indicates that correlations persist even across large temporal separations, and this is referred to as long-range dependence~\citep{beran}. 

If there is power law decay of the autocorrelation function, namely:
\begin{equation}\label{eq:powerlaw}
R_{ss}(\tau) \sim C \tau ^{-\alpha}
\end{equation}
where $C > 0$ and $\alpha \in (0,1)$ are constants, and the symbol $\sim$ indicates asymptotic equivalence~\citep{clegg}, then the time series is said to contain long-range temporal correlations (LRTCs). LRTCs are a subject of considerable scientific interest. They have been detected in biological data~\citep{samorodnitsky, willinger, peng94,carreras, berthouze, linkenkaer01} and have been discussed within the context of complex systems operating in a critical regime.

Applying a Fourier transformation to Equation~\ref{eq:powerlaw}, a similar formulation exists for the spectral density of the signal~\citep{clegg}, with $f$ representing frequency:
\begin{equation}\label{eq:specpower}
G_{ss}(f) \sim B f ^{-\beta}
\end{equation}
where $\beta =1-\alpha$ and is also related to the level of temporal dependence. 

The exponents $\alpha$ and $\beta$ in Equations~\ref{eq:powerlaw} and~\ref{eq:specpower} are connected to the Hurst Exponent, $H$, by $\alpha = 2-2H$ and $\beta = 2H -1$~\citep{taqqu,beran}. 

In practice, finding the exponent $\alpha$ and $\beta$ is not straightforward for an arbitrary signal. In the time-domain, $\alpha$ is best approximated by the slope of the autocorrelation function in the limit of infinite time lags $\tau$ where measurement errors are also largest~\citep{clegg}. Similarly, in the frequency domain, $\beta$ is best approximated by the shape of the spectral density at large frequency shifts $f$. Determination of the Hurst exponent for non-stationary signals is not straightforward, and therefore, for practical applications, the related property of self-similarity (see below) is considered. 

\subsection{Detrended Fluctuation Analysis}\label{sec:DFA}

Detrended fluctuation analysis (DFA) may be used to determine the self-similarity of a time series~\citep{peng94,peng95}. The application of DFA returns the value of an exponent, which is closely related to the Hurst exponent~\citep{clegg,beran}. DFA is often considered to be applicable to both stationary and non-stationary data although recent reports, e.g.,~\citet{bryce}, have suggested that the ability of DFA to deal with non-stationary signals is overstated. In Section~\ref{sec:MLDFA}, we will describe our approach to mitigating this concern.

To calculate the DFA exponent, the time series is first detrended and then cumulatively summed. The root mean square error is then calculated when this signal is fitted by a line over different window sizes (or box sizes). Extensions of the technique can be used to fit any polynomial to each window, however, here we only consider linear detrending. If the time series is self-similar, there will be power law scaling between the residuals (or detrended fluctuations) and the box sizes. In the log space, this power law scaling yields a linear relationship between residuals and box sizes, the so-called DFA fluctuation plot, and the DFA exponent $H$ is obtained using least squares linear regression. A DFA exponent in the range $0.5<H<1$ indicates the presence of long-range temporal correlations. An exponent of  $0<H<0.5$ is obtained when the time series is anti-correlated, $H=1$ represents pink noise, and $H=1.5$ is Brownian noise. Gaussian white noise has an exponent of $H=0.5$.

When performing DFA on oscillatory signals, the smallest window length should be large enough to avoid errors in local root mean square fluctuations, and it is typically taken to be several times the length of a cycle at the characteristic frequency in the time series~\citep{linkenkaer01}. If the minimum window size is significantly smaller than this value, then the fluctuation plot will typically contain a crossover at the window length of a single period~\citep{hu}. However, for non-oscillatory time series for which there is no characteristic temporal scale and there are rapid changes at each innovation, such as Gaussian white noise or FARIMA time series (see Section~\ref{sec:sigsim}), a smaller window size may be used. 

The maximum window size should encompass a significant proportion of the time series yet contain sufficient estimates to allow for a robust estimate of the average fluctuation magnitude across the time series. It is typically taken to be $N/10$ where $N$ is the length of the data~\citep{linkenkaer01}.

In our application of DFA to neurophysiological and model data, we use 20 window sizes with a logarithmic scaling and a minimum window of $8$ time steps for simulated data, and $1$ second for neurophysiological oscillations (sampled at $512$Hz, band-pass filtered $15.5-27.5$Hz) providing for a minimum of $~16$ cycles per second. Following~\citet{linkenkaer01} we take a maximum window size of $N/10$ time steps where $N$ is the length of the time series.

\subsection{Assessing the validity of DFA}
\label{sec:MLDFA}
As mentioned above, a self-similar process will produce a power law relationship between the magnitude of the detrended fluctuations and the box sizes. In DFA, this power law scaling is characterised in terms of the linear scaling between the log detrended fluctuations and the log box sizes (DFA fluctuation plot). It is beyond the scope of this paper to argue the validity of operating in the log domain (but see~\citet{clauset2007} for a reasoned view as to why this may not be appropriate), however, since the object of DFA is to find evidence for or against scaling and because a valid DFA exponent can only be obtained when the DFA fluctuation plot is indeed linear we have introduced a model selection method for establishing the linearity of DFA fluctuation plots~\citep{botcharova13}. 

Our arguments for adopting a more rigourous approach are as follows: i) there is no {\it a priori} means of confirming that a signal is self-similar, ii) a DFA fluctuation plot will necessarily increase with window size, iii) an exponent may be too easily obtained through simple regression analysis producing a statistically significant result with a high $r^2$ value even though the linear model may not best represent a given DFA fluctuation plot, iv) the discovery of an exponent $>0.5$ with a high $r^2$ value may lead to the incorrect conclusion that the signal is self-similar with LRTCs.

Instead of a simple regression we use the model selection technique (ML-DFA) introduced in~\citet{botcharova13} to determine whether a given DFA fluctuation plot is best-approximated by a linear model. This is a heuristic technique, which has been tested extensively and found to perform well in assessing linearity in the fluctuation plots of the following time series: i) those with known combinations of short and long-range temporal correlations, ii) self-similar time series with varying Hurst exponent, iii) self-similar time series with added noise and iv) time series with known oscillatory structure, e.g., sine waves~\citep{botcharova13}. 

The technique fits the DFA fluctuation plot with a number of different models (see below) and compares the fit of each model using the Akaike Information Criterion (AIC), which discounts for the number of parameters needed to fit the model. The DFA exponent is accepted as being valid only if the best fitting model is linear. We want to stress that this does not equate to stating that the fluctuation plot is linear. Rather, we do not reject the linear model hypothesis. In what follows, only those time series for which the linear model hypothesis is not rejected (i.e., their DFA fluctuation plot is best-fitted by the linear model) contribute to the DFA exponents presented in the present paper and where appropriate we indicate where linear scaling of the fluctuation plot is lost.

The models included in ML-DFA are listed below (see~\citet{botcharova13} for a justification), with the $a_i$ parameters to be found. The number of parameters ranges between $2$ for the linear model, and $8$ for the four-segment spline model. 

\begin{list}{}{}
\item Polynomial - $f(x) = \sum_{i=0}^{K} a_i x^i \mbox{ for } K = \left\lbrace 1,...,5 \right\rbrace $
\item Root - $f(x)=a_1 (x + a_2)^{1/K} + a_3 \mbox{ for } K = \left\lbrace 2,3,4\right\rbrace $ 
\item Logarithmic - $f(x)=a_1 \mbox{log}(x + a_2) + a_3$
\item Exponential - $f(x)=a_1 e^{ a_2 x} + a_3$ 
\item Spline with 2, 3 and 4 linear sections.
\end{list}

The first step of ML-DFA is to normalise the fluctuation magnitudes with:
$$lF_{scaled} = 100 \times \frac{lF - lF_{min} }{lF_{max} - lF_{min}} $$
where $lF_{min}$ and $lF_{max}$ are the minimum and the maximum values of vector $lF$ respectively.
A function $\mathcal{L}$ is then defined:
$$ \mathcal{L}  =  \prod_{i=1}^{n} p(lns(i))^{lF_{scaled}(i)} $$ 
which is a product across all windows $i$, and which works in a similar way to a likelihood function, where $p(lns)$ represents the function:
$$p(lns) = \frac{\abs{ f(lns) } }{\sum_{i=1}^{n} \abs{ f(lns) } }$$
where $f(lns)$ is the fitted model. Absolute values are used in order to ensure that $p(lns)$ remains in the range $\left[0, 1\right]   $, so that a function is rejected if it falls below $0$. 

The next step is to apply a logarithm to $\mathcal{L}$ to produce a function that is similar in form to a log-likelihood:
$$ \mbox{log}\mathcal{L}  =  \sum_{i=1}^{n} {lF_{scaled}(i)} \mbox{log} p(lns(i))$$   
This is maximised to find the parameters $a_i$ necessary for $f(lns)$. It is worth mentioning that the application of the logarithm means that the values belonging to $lns$ are not equally weighted for all $i$. The larger window sizes have a lower weighting, which is beneficial because these estimates are also the least robust since they have fewer samples associated with them.

Akaike's Information Criterion (AIC) is then computed, which is designed to prevent over-fitting -- a situation that should in general be avoided -- by taking into account the number of parameters used~\citep{akaike,mackay}. For a model using $k$ parameters, with likelihood function $ \mbox{log}\mathcal{L} $, the Akaike Information Criterion is calculated using the following expression:
$$ \mbox{AIC} = 2k - 2 \mbox{log}\mathcal{L} + \frac{2k(k+1)}{n-k-1}$$ 
where $k$ is the number of parameters that the model uses~\citep{akaike}. An adapted formula was proposed by~\citet{hurvich}, which accounts for small sample sizes. The model which provides the best fit to the data is that with the lowest value of AIC. It is important to recall that the AIC can only be used to compare models. It does not give any information as to how good the models are at fitting the data, i.e., it is only its relative value, for different models, that is important; and it would not be possible, for instance, to compare AIC values obtained from different data sets to each other.

\subsection{Method Validation}

\subsubsection{FARIMA processes.}\label{sec:sigsim}
An Autoregressive Fractionally Integrated Moving Average model (FARIMA)~\citep{hosking} can be used to create time series with self-similarity. The model provides a process that can easily be manipulated to include a variable level of LRTCs within a signal, from which DFA should return the exponent used to construct the FARIMA process.

To construct a FARIMA process a time sequence of zero-mean white noise is first generated, which is typically taken to be Gaussian, and necessarily so to produce fractional Gaussian noise. The FARIMA process, $X(t)$, is then defined by parameters $p$, $d$ and $q$ and given by:

\begin{equation}\label{eq:farima}
\left(1-\sum_{i=1}^{p}{\varphi_i B^i} \right) \left( 1-B\right)^d X(t)= \left(1+\sum_{i=1}^{q}{\varphi_i B^i} \right)  \varepsilon(t).
\end{equation}

$B$ is the backshift operator operator, so that $B X(t) = X(t-1)$ and $B^2 X(t) = X(t-2)$. Terms such as $(1-B)^2$ are calculated using ordinary expansion, so that $(1-B)^2 X(t) = X(t) - 2 X(t-1) + X(t-2)$. While the parameter $d$ must be an integer in the ARIMA model, the FARIMA can take fractional values for $d$. A binomial series expansion is used to calculate the result:

$$ (1-B)^d = \sum_{k=0}^{\infty} \binom{d}{k} (-B)^k.$$

The left hand sum deals with the autoregressive part of the model where $p$ indicates the number of back-shifted terms of $X(t)$ to be included, $\varphi_i$ are the coefficients with which these terms are weighted. The right hand sum represents the moving average part of the model. The number of terms of white noise to be included are $q$, with coefficients $\varphi_i$. In the range $\vert d \vert < \frac{1}{2}$, FARIMA processes are capable of modelling long-term persistence~\citep{hosking}. As we will only consider $p=1$ and $q=1$ throughout the manuscript, we will refer to $\varphi_1$ as $\varphi$ and $\varphi_1$ as $\theta$. We set $\vert \varphi \vert < 1$, $\vert \theta \vert < 1$ to ensure that the coefficients in Equation \ref{eq:farima} decrease with increasing application of the backshift operator, thereby guaranteeing that the series converges, and $X(t)$ is finite~\citep{hosking}.

A FARIMA(0,$d$,0) is equivalent to fractional Gaussian noise with $d=H-\frac{1}{2}$ ~\citep{hosking}. This produces a time series with a DFA fluctuation plot that has been shown to be asymptotically linear with a slope of $d+0.5$~\citep{taqqu,bardet}. By manipulating the $\varphi$ and $\theta$ parameters, the DFA fluctuation plots can also be distorted.

\subsubsection{Surrogate Data.}\label{sec:surrogatedata}

Two time series $x_1(t)$ and $x_2(t)$ can be constructed such that the time derivative of their phase difference is a FARIMA time series $X(t)$ with a known DFA exponent~\citep{hosking}. Concretely, we work backwards from the time series $X(t)$ to which DFA is applied. The phase difference of the two time series $\Delta(\phi(t)$ will be the cumulative sum of $X(t)$, which is discrete in this case: 
$$\Delta(\phi(t)) = \sum_{s=1}^{t} X(s)$$

The two phases $\phi_i(t)$ and $\phi_2(t)$ of $x_1(t)$ and $x_2(t)$ respectively must be constructed to have a difference of $\Delta(\phi(t))$, or some multiple of $\Delta(\phi(t))$ since DFA is unaffected by multiplying a time series by a constant. We therefore set $\phi_1(t) = \frac{\sum_{s=1}^{t} X(s)}{2 f_s}$ and $\phi_2(t) = -\frac{\sum_{s=1}^{t} X(s)}{2 f_s}$ where $f_s$ takes the role of a nominal sampling rate for the surrogate data.  

Since the phase of a cosine signal is equal to its argument, the two signals $x_1(t)$ and $x_2(t)$ are defined as: 
$$  x_1 = \mbox{cos}(\omega + \frac{\sum_{s=1}^{t} X(s)}{2 f_s}) $$
and
$$ x_2 = \mbox{cos}(\omega - \frac{\sum_{s=1}^{t} X(s)}{2 f_s}) $$
where $\omega$ is a constant.

In what follows, we used $\omega=1$ and $f_s=600$. These values were chosen in order to produce a smooth enough phase difference. This was necessary to prevent artefacts produced by the Hilbert transform when applied to non-smooth data. When using physiological data, a high enough sampling rate guarantees that the signals will be smooth. 

A hundred time series $X(t)$ were generated using the algorithm described in ~\citep{hosking} for each of the $11$ DFA exponents $0.5, 0.55,0.6,...,1$. Each simulation contains $2^{22} = 4194304$ innovations. The value of the exponent of $X(t)$ is first computed, the two signals $x_1(t)$ and $x_2(t)$ are then constructed, and the phase analysis method is applied. Window sizes used for application of DFA were logarithmically spaced with a minimum of $600$ time steps to correspond to $f_s$ and maximum $N/10$ where $N = 2^{22}$ is the length of the time series. 

A further control analysis was performed in which a Gaussian white noise time series $\eta_i(t) $ was added to one of the signals, namely, 
$$x'_1(t) = \mbox{cos}(\omega + \frac{\sum_{s=1}^{t} X(s)}{2}) + \eta_i(t)$$
before the phase analysis method was applied in order to recover the DFA exponent of the phase difference $X(t)$. This allowed us to alter the signal-to-noise ratio of $x_1(t)$ in an additive way, which we may suppose to be the case for noise in a neurophysiological time series. By applying the phase analysis method to signals with additive noise, we were able to test the robustness of the method to noisy data. In this analysis, first we will estimate the extent to which the DFA exponent alters when noise is added. Second, we will assess whether ML-DFA rejects those DFA exponents that we know to contain noise, and if so, we will quantify the level of noise at which exponents are no longer valid.

\subsection{Model Simulations}

\subsubsection{The Ising model.}\label{sec:ising}
The Ising model is a model of ferromagnetism~\citep{ising}. In two dimensions, the model is implemented on a lattice (grid) of elements, or particles which represent a metallic sheet. A temperature parameter controls the collective magnetisation~\citep{onsager}. The Ising model has been recently used as a model for a 2-dimensional network of connected and interacting neurons~\citep{bullmore}. 

Each element of the grid is assigned a spin $p_i$, initially at random, which takes a value $+1$ (spin up) or $-1$ (spin down). Spins may switch up and down in time in a fashion influenced by both the energy of the full system and by the spin configuration of other neighbouring elements. The energy of the system in a given configuration of spins $p$ is given by the Hamiltonian function $H(p) = -J \Sigma^{N}_{i,j = nn(i)} p_i p_j$, where $j$ is an index for the four elements that are nearest neighbours $nn$ of each element, $i$ of the square grid. The negative sign is included by convention. The average energy of the system $E = <H>$ where the symbol $<>$ indicates taking the expectation value.

The probability $\mathbb{P}$ of a given configuration occurring is then proportional to $\mathbb{P} = e^{-H(p)/kT}$, where $T$ is the temperature parameter and $k$ is Boltzmann's constant. The system may switch into a new configuration if its associated probability is higher or equal to that of the current configuration. The Ising model is implemented using the Metropolis Monte Carlo Algorithm~\citep{metropolis}. 

\begin{figure}
\begin{center}
\includegraphics[width=15cm]{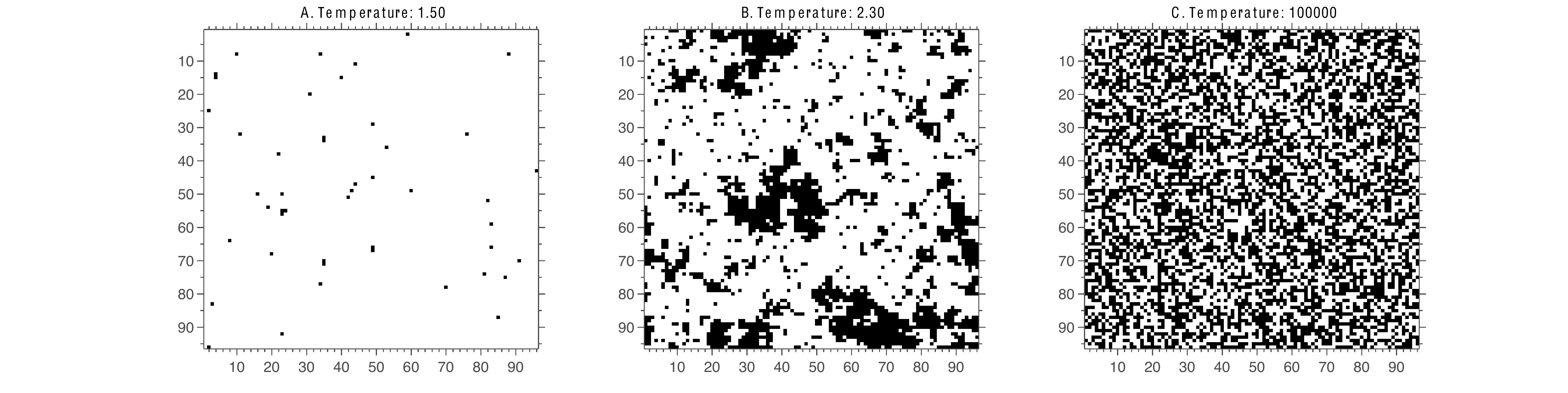}
\end{center}
 \textbf{\refstepcounter{figure}\label{fig:isinggrid} Figure \arabic{figure}. The Ising model lattice at a single time point once steady state has been reached for 3 different values of the temperature parameter.}{ A. The Ising lattice at a cold temperature of $1.5$. Almost all spins are aligned (white) and there is little change across time. C. The Ising lattice at a high temperature of $T=10^5$. The spins form a more or less random pattern across the lattice. B. The Ising lattice near critical temperature, $T=2.3$. The lattice contains clusters of spins that are both small and large. Note that these are snapshots and that the spin structure of the model is best appreciated when evolving across time.}
\end{figure}

At temperature $T=0$, the system is highly ordered and corresponds to a magnetic state (see Figure~\ref{fig:isinggrid} for an example of an Ising model lattice). With increasing temperature values, the probability of a spin changing increases. As the system temperature increases the spins change more rapidly and the system becomes increasingly disordered and corresponds to a non-magnetic state (Figure~\ref{fig:isinggrid}A). The temperature value at which a transition occurs between the magnetised and non-magnetised states is known as the critical temperature $T_c$. At this temperature (see Figure~\ref{fig:isinggrid}B), the system will have a large dynamic range and infinite correlation length. However, in practice, this means that the system contains spin clusters of all sizes, and correlations between elements of an infinite system remain finite~\citep{onsager,daido}. In other words, the Ising model is predicted to have long-range correlations between its elements at $T_c$. 

The value of the critical temperature $T_c$ was calculated for the 2-dimensional Ising model in \cite{onsager}, and is given by the solution to the equation $$\mbox{sinh}\left( \frac{2J}{k T_c} \right) = 1 $$


In the implementation of the Ising model used here, the lattice consists of $96 \times 96$ elements. The constants $J$ and $k$ are set to $J=1$ and $k=1$ without loss of generality, which gives the critical temperature $T_c = \frac{2}{ln(1+\sqrt{2})} \approx 2.269$.

In order to obtain a time series from this spatial model, we follow the procedure introduced by~\citet{bullmore}. Namely, the lattice is divided into a number of smaller square lattices, which we refer to as sub-lattices, and a number of time series are created by taking an average spin value for each sub-lattice. Here, we use a sub-lattice size of $8 \times 8$ as in~\citet{bullmore}, but we also investigated other sub-lattice sizes (results not shown) in order to verify that this choice of sub-lattice size did not affect the results. Indeed, previous work by~\citet{viola} suggests that the sub-sampling of a system may cause it to be mis-classified as sub-critical or supercritical when it is in fact in a critical state.

Pairs of time series, for every possible pairing of sub-lattices belonging to the larger grid, were used as input signals for the phase analysis method. For the sub-lattice of size $8 \times 8$ considered here, $144$ time series could be created allowing for $10,296$ pairings. 

\subsubsection{The Kuramoto model.}\label{sec:kuramoto}

The Kuramoto model is a classical model of synchronisation~\citep{acebron,chopra05} and has been used to study the oscillatory behaviour of neuronal firing~\citep{breakspear,bullmore,pikovsky} among many other biological systems. 

The Kuramoto model describes the phase behaviour of a system of mutually coupled oscillators with a set of differential equations. Each of $N$ oscillators in the system rotates at its own natural frequency $\left\lbrace \omega_i, i = 1,...,N \right\rbrace $, drawn from some distribution $g(\omega)$. However, it is attracted out of this cycle through coupling $K$, which is globally applied to the system. Time $t$ is taken to run for $T$ seconds of length $dt = 10^{-3}$. 
The differential equation to describe the phase of an oscillator is~\citep{kuramoto1975,kuramotobook}:

\begin{equation}\label{eq:thetadot}
\dot{\phi}_i(t) = \omega_i(t) + \frac{K}{N} \Sigma^{N}_{j = 1} \mbox{sin}(\phi_j (t)- \phi_i(t)) 
\end{equation}

Because the Kuramoto model provides an equation governing the phase evolution of each oscillator in the system, there is no need for the Hilbert transform to recover the phase time series and therefore only the latter stages of the phase analysis method are used (see steps 3-6 in Figure \ref{fig:method}).  

\citet{kuramoto1975} showed that the evolution of any phase $\phi_i (t)$ may be re-expressed using two mean field parameters, which result from the combined effect of all oscillators in the system. Namely, we may write:

\begin{equation}\label{eq:meanfield}
\dot{\phi}_i(t) = \omega_i + K r(t) \mbox{sin}(\psi(t) - \phi_i(t)) 
\end{equation}

where $\psi(t)$ is the mean phase of the oscillators, and $r(t)$ is their phase coherence, so that:

\begin{equation}\label{eq:order}
r(t) e^{i \psi(t)} = \frac{1}{N} \sum^N_{j=1} e^{i \phi_j(t)}
\end{equation}

This crucially indicates that each oscillator is coupled to the others through its relationship with mean field parameters $r(t)$ and $\psi(t)$, so that no single oscillator, or oscillator pair drives the process on their own. The oscillators synchronise at a phase equal to the mean field $\psi(t)$, and $r(t)$ describes the strength of synchronisation, sometimes referred to as the extent of order in the system~\citep{mirollo,bonilla}. When $r(t)=0$, no oscillators are synchronised with each other. When $r(t)=1$, all oscillators are entrained with each other.

One solution to Equation \ref{eq:meanfield} is $r \equiv 0$ for all time and coupling, leaving each oscillator to evolve independently at its own natural frequency. Using a limit of $N \rightarrow \infty$, some further deductions can be made, including the fact that when the natural frequency distribution $g(\omega)$ is unimodal and symmetric, another solution can be found for $\omega_i$, with $r(t)$ not equivalent to $0$~\citep{kuramoto1975}. A critical bifurcation occurs for sufficiently high coupling, resembling a second-order phase transition~\citep{miritello} in which the order parameter (here, $r(t)$) leaves zero and grows continuously with coupling~\citep{dorfler, mirollo}. The coupling at the bifurcation is referred to as the critical coupling $K_c$~\citep{dorfler}. 

In an infinite Kuramoto model, criticality is defined through this point of bifurcation. For a finite system, however, the critical point can only be approximated by this theoretical value. One defining characteristic of the critical coupling for the Kuramoto system is that the greatest number of oscillators come into synchronisation at this value. In our study, we deal with finite-sized implementations of the Kuramoto model, and we use this characteristic as a marker of the onset of critical regime in addition to the theoretical value $K_c$. Specifically, we use a measure characterising the onset of synchronisation with increasing coupling introduced by~\citet{bullmore}. This is the change in the `effective mean-field coupling strength', $\Delta (Kr)$. If the value of $Kr$ exceeds the difference between the natural frequency and the mean phase $\omega_i - \psi$ (in modulus), i.e., $\vert \omega_i - \psi \vert <  Kr$, then oscillator $i$ will synchronise to the mean field ~\citep{mertens}. Thus the value of $K$ at which $Kr$ increases maximally is the coupling value at which the greatest number of oscillators are drawn into the mean field.


In this paper, we consider the Kuramoto model with a noise term added to the phase equation, namely, Equation \ref{eq:thetadot} becomes:
\begin{equation}\label{eq:thetadotnoise}
\dot{\phi}_i(t) = \omega_i(t) + \frac{K}{N} \sum^{N}_{j = 1} \mbox{sin}(\phi_j (t)- \phi_i(t)) + \eta_i(t) 
\end{equation}

where $\eta_i$ is a noise input taken to be uncorrelated Gaussian noise with zero mean ($\left\langle \eta_i \right\rangle = 0$) and covariance $\sigma_i^2/T$  ($\left\langle \eta_i (t)\right\rangle \left\langle \eta_j (s)\right\rangle = \delta_{ij}\delta(t-s) \sigma_i^2/T$) where $\delta_{ij}$ is the Kronecker delta, $\delta(t-s)$ is the Dirac delta function, $\sigma_i$ is in radians and $T=1$ second here. 

This creates a richer structure in the oscillator dynamics, which we suggest may better reflect coupling of neurophysiological oscillators. Furthermore, it has been shown that addition of noise increases the critical regime over a wider range of coupling values~\citep{breakspear}. This may allow for the fluctuations of phase difference of a given oscillator pair to persist for longer with increasing coupling before full synchronisation is achieved. 

In~\citet{daniels}, the critical coupling for the infinite Kuramoto model with added noise $K_{c,noise}$ is calculated to be:

$$K_{c,noise} = 2 \left[ \int_{-\infty}^{\infty} \frac{\sigma_{\omega}}{\sigma_{\omega}^2 + \omega^2} g(\omega) d \omega \right]^{-1} $$

Again, as the number of oscillators is inevitably finite, this value is only an approximation to the true critical coupling in the system, but we find it useful and it is displayed alongside plots of $\Delta (Kr)$, which although originally introduced for a noiseless model, remains a helpful marker of the effective critical coupling in the Kuramoto model when noise levels are not too large~\citep{mertens}.

In this study, we generated time series for $200$ oscillators of the Kuramoto model described by Equation~\ref{eq:thetadotnoise}. Each time series was 6100-timestep long. The standard deviation $\sigma_i$ was set to $0.32$. The distribution of natural frequencies was $g(\omega) \sim \mathcal{N} (44 \pi,\sigma_{\omega} ) $, with standard deviation $\sigma_{\omega} = 15$. This corresponds to a normal distribution centred around $22$ Hz (which is a unimodal distribution). In order to get an idea of the spread of the distribution, the minimum natural frequency selected from this distribution was $16.3$ Hz and the maximum was $27.8$ Hz. We selected this frequency range because it spans the $\beta$-band of EEG, MEG and EMG oscillations~\citep{farmer98}.

For these parameter values, the critical coupling $K_c$ is equal to:
$$K_c = \frac{2 \sqrt{2}}{\sqrt{\pi}} \sigma_{\omega} \sim 23.93$$
The integral for $K_{c,noise}$ is not analytically calculable for a normal distribution $g(\omega) \sim \mathcal{N}$, but empirical calculation yields:
$$K_{c,noise} \sim 23.85 $$

\subsubsection{The Cabral model.}\label{sec:cabral}

The third model that we consider in this paper was developed by Joanna Cabral and her colleagues, referred to as the Cabral model. It is a modification of the Kuramoto model, combining the dynamics of the Kuramoto oscillators with the network properties observed in the human brain~\citep{cabral}.

\begin{figure}
\begin{center}
\includegraphics[width=10cm]{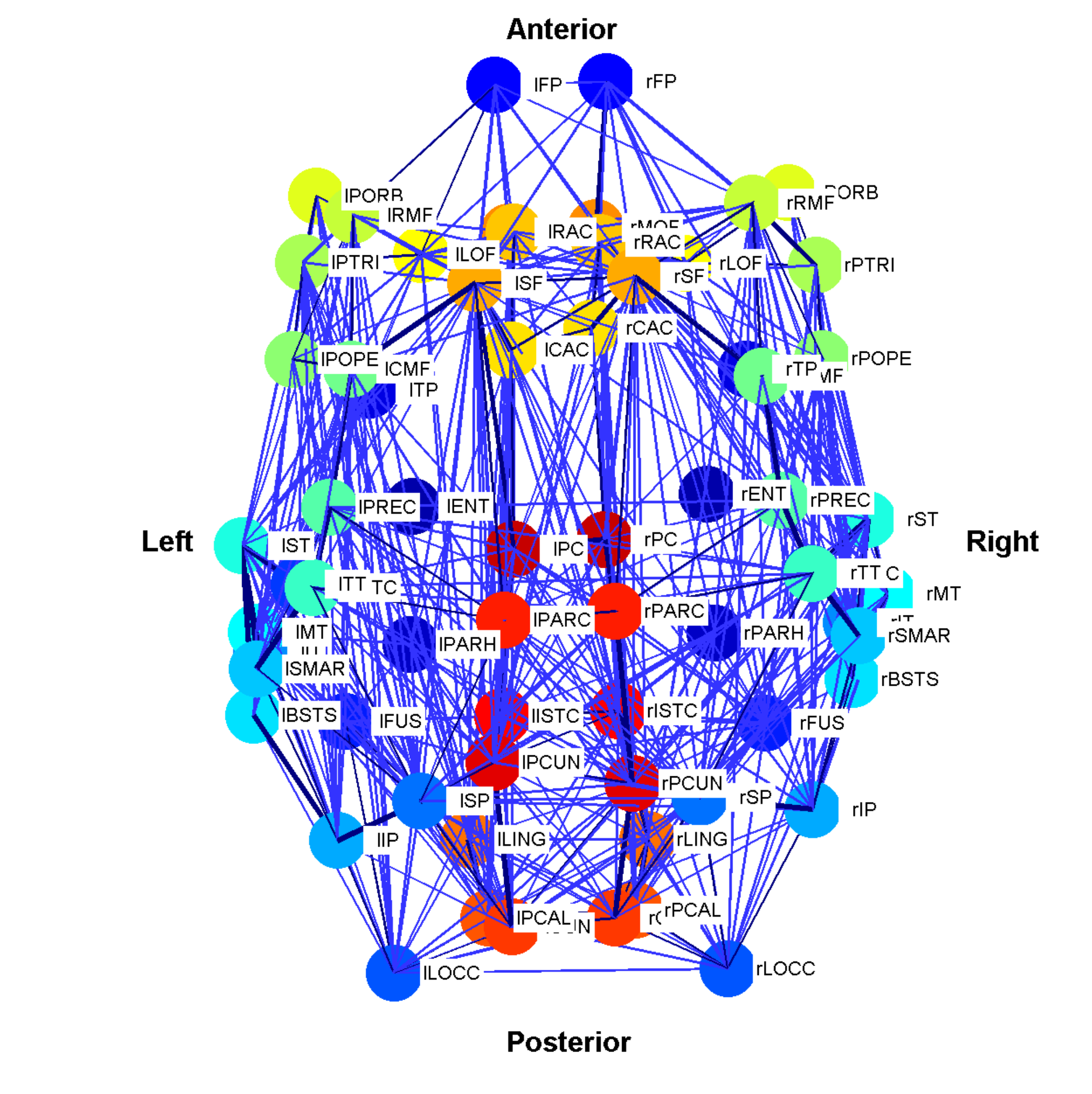}
\end{center}
 \textbf{\refstepcounter{figure}\label{fig:cabralBrain} Figure~\arabic{figure}. Schematic plot (top view) of the Cabral human brain model showing the connections and connection weights between oscillators which correspond to different brain regions.}{ The weight of the connection lines represent the strength of connectivity between the oscillators. The darkest blue lines are the strongest 1\% of connections. The node colours represent oscillators, which model different brain regions as detailed in~\citet{cabral}. Colours are consistent for homologous regions in the left and right hemispheres. Anterior and posterior, left and right are shown.}
\end{figure}

The Cabral model includes a noise input to the Kuramoto oscillators and situates the 66 oscillators on a connectivity matrix with varying connection strengths and time delays based on empirical measurements of 998 brain regions, which have been down-sampled to 66~\citep{honey09}. The list of brain regions considered in this model are given in the supplementary material of~\citet{cabral} and are reproduced in the Appendix to the present paper. Specifically, Equation~\ref{eq:thetadot} is modified to include a connectivity term $C_{ij}$ between oscillators $j$ and $i$, namely, 
\begin{equation}\label{eq:cabral}
\dot{\phi}_i(t) = \omega_i(t) + \frac{K}{N} \Sigma^{N}_{j = 1} C_{ij}\mbox{sin}(\phi_j (t-D_{ij})- \phi_i(t)) + \eta_i(t) 
\end{equation}
where $\eta_i$ is the noise input previously introduced, and $D_{ij}$ is the time delay associated with the link between oscillators $j$ and $i$. The matrix of delays $D$ is extracted from a matrix of empirical distances $L$ between regions using:
$$D_{ij} = \frac{\left\langle D \right\rangle L_{ij}}{\left\langle L \right\rangle}  $$
and is used to encode the length of time taken by neural activity to traverse the connection space. The connectivity and distance matrices ($C$ and $D$, respectively) can be visualised through the schematic diagram shown in Figure~\ref{fig:cabralBrain}. The thickness and colour of the lines in the diagram represent the weights of the connections between the oscillators representing individual brain regions. These weights are proportional to the number of fibres that were empirically observed to connect the various regions~\citep{cabral,cabral2012}. Brain regions may be identified by their labels, the abbreviations of which are given in Table~\ref{tab:brains} in the Appendix. 

In~\citet{cabral}, the model was used to generate time series which were used as input to a hemodynamic model and bandpass filtered. The resulting time series were compared to recordings of BOLD fMRI signals using Pearson's correlation coefficient and mean squared error to determine the parameter values $K$ and $\left\langle D \right\rangle $ that generated the time series which most closely approximated the BOLD data. 

In this model, there is no theoretically derived value of critical coupling and $\Delta (Kr)$ is only a marker of effective change in coupling that may or may not be critical. We interpret a rise in $\Delta (Kr)$ as an increase in order of the system similar to that observed by~\citet{bullmore}.


The phase analysis method presented here was applied to the Cabral model for coupling parameters $K$ ranging from $1$ to $20$. We note that this encompasses $K=18$, the value identified by~\citet{cabral} as best approximating human brain resting state BOLD fluctuations. Natural frequencies were drawn from a normal distribution with $g(\omega) \sim \mathcal{N} (120 \pi,\sigma_{\omega} )$ with standard deviation $\sigma_{\omega} = 5$, which corresponds to a normal distribution centred around $60$ Hz in the $\gamma$ frequency band. This was selected because $\gamma$ oscillations have been shown to play a significant part in the BOLD signal fluctuations (see~\citet{cabral} for details).

The standard deviation $\sigma_i$ of the noise input was set to $1.25$. It was found that values of $\sigma_i<3$ did not significantly alter the resulting parameter values of $K$ and $\left\langle D \right\rangle $. The value $\left\langle D \right\rangle = 11$ is taken as in~\citet{cabral}.  


\subsubsection{Clusters in the Cabral model.}
\citet{cabral} identified a number of clusters of oscillators, along with a set of 12 oscillators which are not part of a cluster. These clusters are listed below in Table~\ref{tab:clusters}. In our analysis, we considered how each of these different clusters contributed to the overall behaviour. 
 
\begin{table}[ht]
\caption{\bf{Cluster information.} The 66 oscillators of the Cabral model can be separated into 6 clusters, based on their mutual connectivity and distance matrix patterns, and a final set of 12 oscillators, which are not considered to belong to a cluster, but are grouped together here for convenience. The table also states the average sum of weights per node belonging to each cluster and the average number of connections per node (both to 2 d.p.). } 
\centering 
\begin{tabular}{|c|c|c|c|} 
\hline 
Clusters & Oscillators & Average weight per node & Average degree distribution \\
 [0.5ex] 
\hline 
Cluster 1	&	7-17 & 0.29 & 19.09 \\
Cluster 2	&	18-22	& 0.16 &  15.80 \\
Cluster 3	&	23-26,41-44	& 0.30 &  21.00 \\
Cluster 4	&	27-40	& 0.34 &  21.71 \\
Cluster 5	&	45-49	& 0.15 & 	 15.60   \\
Cluster 6	&	50-60	& 0.27 & 18.73  \\
Individual Oscillators	&	1-6,61-66 & 0.03 &  08.59	  \\
\hline 
\end{tabular}
\begin{flushleft}    
\end{flushleft}
\label{tab:clusters} 
\end{table}

\subsubsection{Disruptions to the Cabral model.}\label{sec:cabraldisrupted}

In order to investigate the role of connectivity in sustaining LRTCs of rate of change of phase difference, we modified the connectivity matrix $C$ in the Cabral model in two ways. First, beginning with the empirical connectivity matrix we deleted any connection that extended from one hemisphere into the other. We preserved all the other elements of the model's connectivity and oscillator characteristics. 

\begin{figure}
\begin{center}
\includegraphics[width=\textwidth]{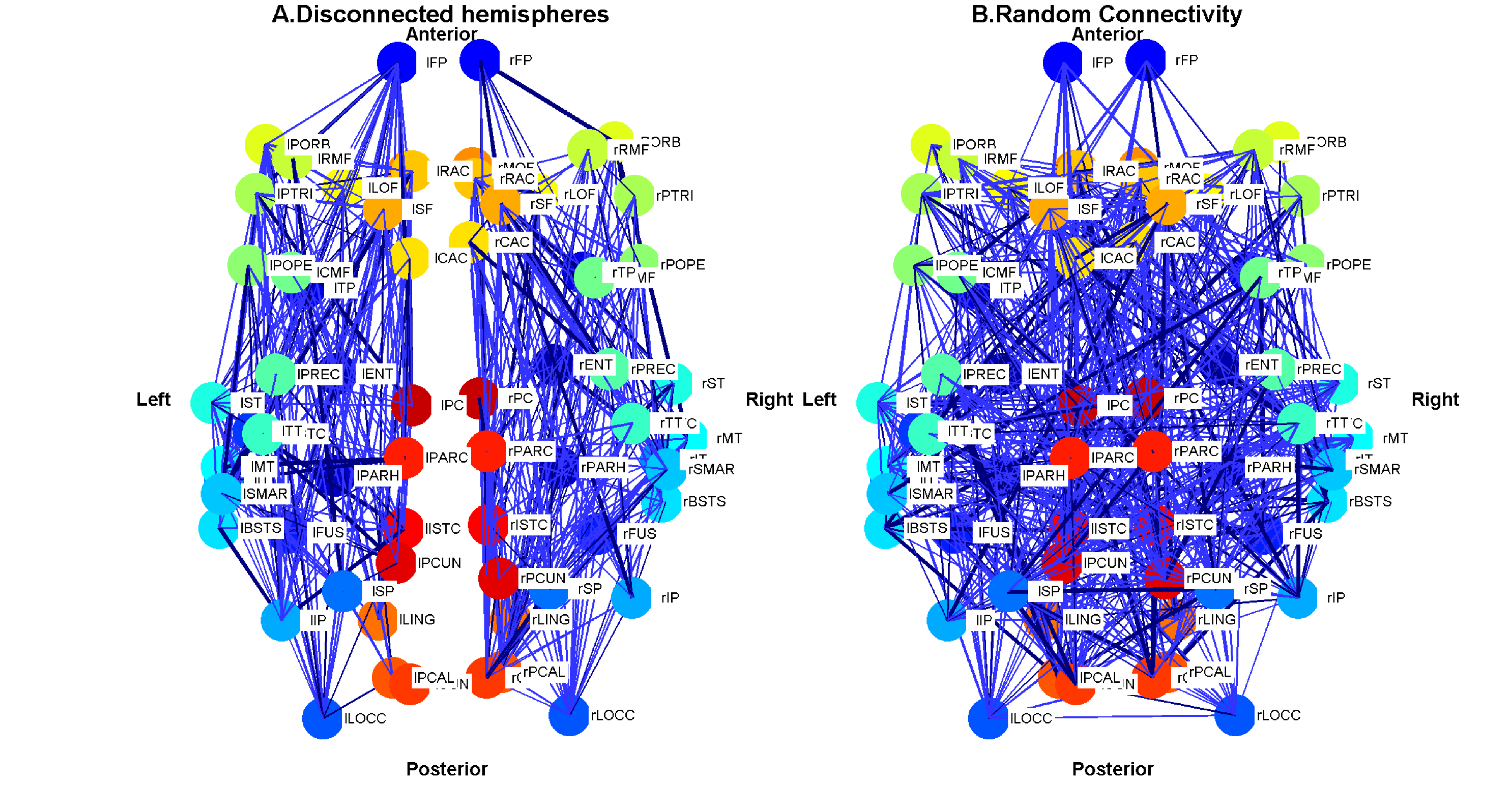}
\end{center}
 \textbf{\refstepcounter{figure}\label{fig:disconnectBrain} Figure \arabic{figure}. Schematic plot (top view) showing the connections and connection weights between oscillators belonging to two modifications to the connectivity of the Cabral human brain model.}{ A. The left and right hemispheres of the brain have been disconnected, but connections within each hemisphere are left unchanged. B. The connections and weights of each node are assigned randomly, but the degree distribution and weight distribution at each node is kept constant. The weight of the connection lines represent the strength of connectivity between the oscillators. The darkest blue lines are the strongest 1\% of connections. The node colours represent oscillators, which model different brain regions as detailed in~\citet{cabral} and are identical to Figure~\ref{fig:cabralBrain}. Colours are consistent for homologous regions in the left and right hemispheres. Anterior and posterior, left and right are shown.}
\end{figure} 

The second exploration involved a reconnection of the connectivity matrix in a random arrangement, while preserving the degree distribution and weight distribution of each oscillator by an algorithm described in~\citet{gionis, hanhijarvi}. Specifically, a list of the outgoing weights of each oscillator was made alongside the node from which it extends. Two weights were selected from this list. If they did not belong to the same node, then the nodes were connected to each other with the associated outgoing weights that were selected. These weights were then deleted from the list. To continue the algorithm, two further weights were selected. After the first step, it was necessary to check at each iteration that the nodes were not already connected before connecting them. If the nodes were connected, or if they were the same node, new weights were selected from the list. 

Analysis of the random connectivity model and comparison of the results obtained from it to those derived from the disconnected hemisphere model and standard appropriately connected model allowed us to determine the extent to which a realistic connectivity matrix of the human brain predisposes the system to LRTCs in the rate of change of the phase difference between the oscillator pairs representing different brain regions.

 
\paragraph*{A note on notation.}

From this point in the text, all instances of oscillator phase $\phi_i(t)$ and $r(t)$ will be written as $\phi_i$ and $r$ for ease of notation, unless stated otherwise. Any quantities that are defined using the phases of one or more oscillators are also implicitly functions of time, although the $t$ is omitted for the same reason.

\subsection{Neurophysiological data}\label{sec:physiodata}
Previously collected neurophysiological data were used to illustrate the application of the method (see ~\citep{james} for full details). Briefly, EEG and EMG signals were simultaneously recorded whilst a healthy adult subject performed a 2-minute $10\%$ MVC (maximum voluntary contraction) isometric abduction of the index finger of the right hand.  The EMG was recorded using bipolar electrodes situated over the first dorsal interosseous muscle (1DI).  The EEG was recorded using a modified Maudsley montage from 24 Ag/AgCl electrodes with impedance $<5k\Omega$.  The data were amplified and bandpass filtered $4-256$Hz and sampled at 512Hz. We analysed EEG recorded from over the left sensorimotor cortex. The signal processing pathway was set out as in Figure~\ref{fig:method}, including bandpass filtering in the $\beta$ frequency range ($15.5-27.5$Hz). 

\section{Results}

\subsection{Surrogate Data}

\begin{figure}
\begin{center}
\includegraphics[scale=0.4]{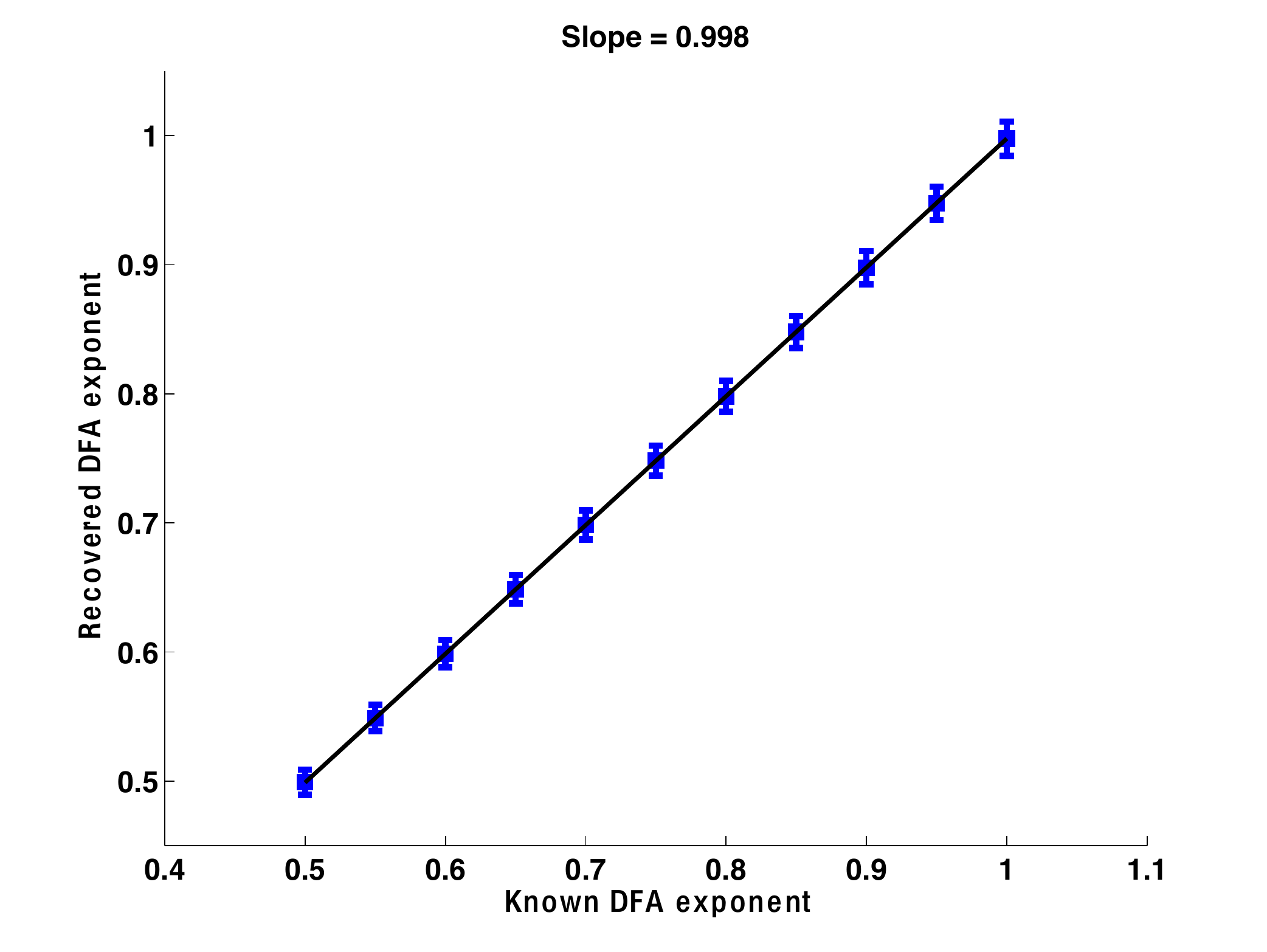}
\end{center}
\textbf{\refstepcounter{figure}\label{fig:surrogate} Figure \arabic{figure}. Plot of the recovered against the true DFA exponent for FARIMA time series.}{ The relationship between recovered and true DFA values is well-approximated by a linear trend with a slope of almost 1. The error bars increase very slightly with increasing DFA exponent.}
\end{figure}

The signals described in Section~\ref{sec:surrogatedata} were analysed. The scatter plot presented in Figure~\ref{fig:surrogate}A shows the DFA exponents of the rate of change of phase difference expected from the construction of a FARIMA time series with known parameters against those recovered by applying the phase analysis method. The scatter plot shows a strong linear relationship between the expected and recovered exponents with a slope of 0.998. The fact that the slope is slightly $<1$ indicates that the recovered exponent was slightly under-estimated by our method. This minor tendency will decrease the likelihood of false positive results.

As noise is added to a signal with a known DFA exponent in its phase, the exponent of its phase is found to be reduced. Figure~\ref{fig:surrogate}B shows that as the noise level is progressively increased, the percentage difference between the known DFA exponent and that recovered by the method increases. When the noise level is above one which causes the percentage difference between known and recovered DFA exponent to exceed approximately 5\% (note, as shown in Figure~\ref{fig:surrogate}B, that this noise level depends on the exponent, e.g., $~0.1$ for true DFA exponent of $1$, $~0.025$ for exponent of $0.75$), no values are returned for the recovered DFA exponent. This occurs because the recovered DFA exponents are not considered to be valid by ML-DFA because their associated DFA fluctuation plots are not best approximated by a linear model (see Section~\ref{sec:MLDFA}). 

As the noise level is increased further, and as it passes a level of $\approx 0.3-0.4$, noise dominates the signal and valid exponents are once again obtained. These exponents are at or close to 0.5 regardless of the value of the known DFA exponents, indicating that the phase relationship of the two signals $s_1(t)$ and $s_2(t)$ is dominated by noise only.

\begin{figure}
\begin{center}
\includegraphics[scale=0.40]{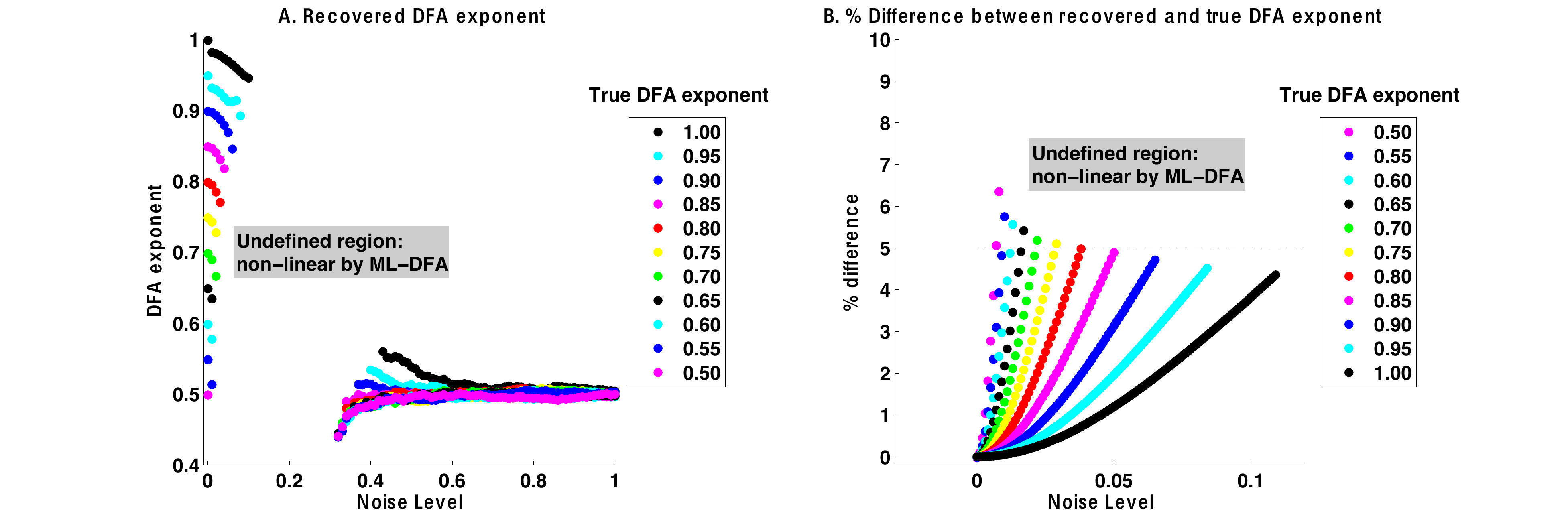}
\end{center}
\textbf{\refstepcounter{figure}\label{fig:surrogate5} Figure \arabic{figure}. True and recovered DFA exponents for noisy signals with LRTCs.}{ A. Recovered DFA exponent values as noise is progressively added. For each of the DFA exponents given in the legend (box insert), a signal $x'_1(t)$ was constructed with a noise level $\sigma \in [0,1]$, shown on the $x$ axis. The phase synchrony analysis method was applied to $x'_1(t)$ and $x_2(t)$. This was performed 100 times. For DFA exponents corresponding to DFA fluctuation plots that were accepted as linear by ML-DFA, the average value for the 100 signal pairs is shown. There are no data points corresponding to the intermediate noise level of $\approx 0.1$ to $\approx 0.3$ because all 100 DFA fluctuation plots for signals with this noise level were determined to be invalid by ML-DFA. B. The \% difference between recovered and known DFA exponents as a function of the noise added to a signal with a known DFA exponent in its phase. The data shown in this plot is the same as that in panel A, but it is expressed in terms of the \% difference between true and recovered DFA exponents rather than the raw recovered value. Only noise levels of $\sigma \in [0,  0.1]$ are shown. The colours represent different true DFA exponent values, as indicated by the legend within the inserted box. The dashed line indicates a 5\% difference between known and recovered exponents. When the difference between the known and recovered exponent exceeded approximately 5\% for any value of the true exponent, the DFA fluctuation plot is not accepted as being linear by ML-DFA and therefore the exponent is not shown on the plots.}
\end{figure}

\subsection{The Ising model}
\label{sec:isingresults}
\begin{figure}
\begin{center}
\includegraphics[width=15cm]{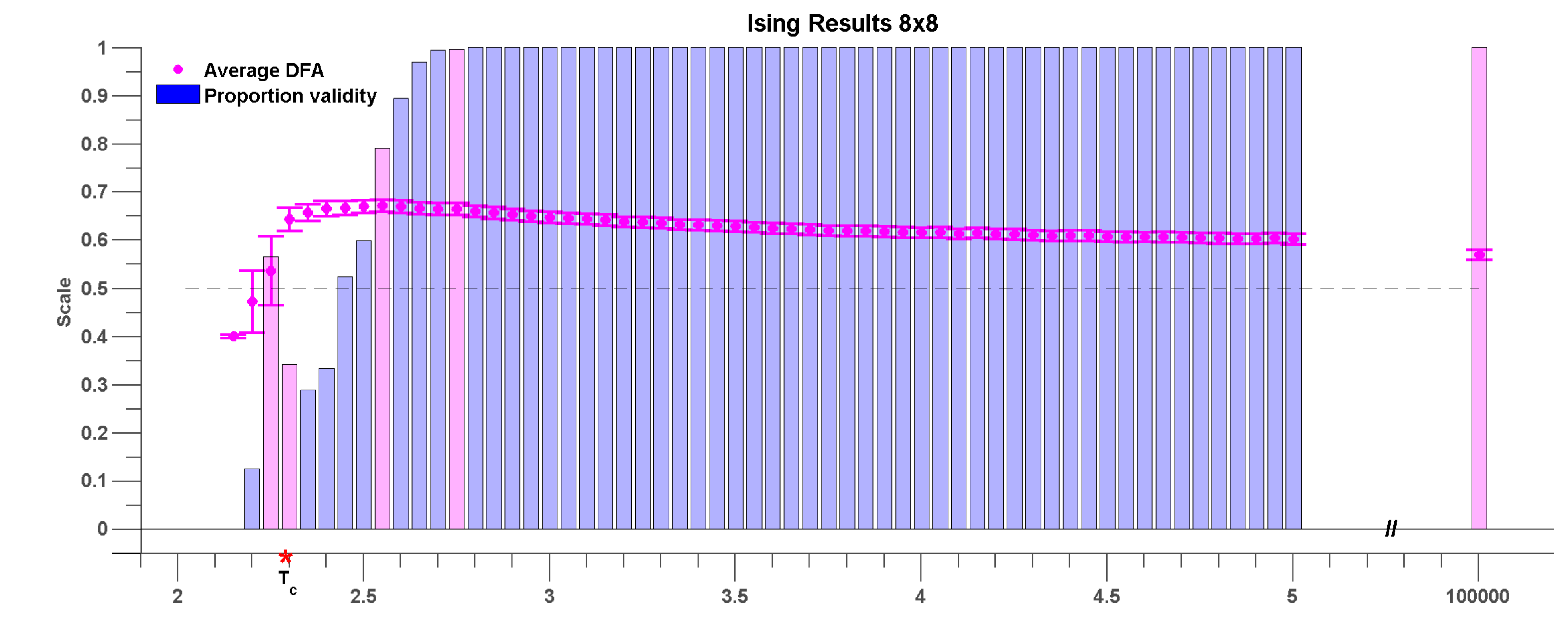}
\end{center}
 \textbf{\refstepcounter{figure}\label{fig:ising6} Figure \arabic{figure}. Average DFA exponents of rate of change of phase difference between pairs of time series generated by $8 \times 8$ sub-lattices of the $96 \times 96$ Ising model lattice.}{ The temperature parameter, $T$, is varied on the $x$ axis. The average of the valid DFA exponents is shown in pink, and the error bars are a single standard deviation from the mean. The proportion of valid exponents, as calculated by ML-DFA, is denoted by the vertical bars. The theoretical critical parameter $T_c$ is indicated by a red asterisk. A horizontal dashed line at DFA exponent 0.5 is plotted to guide the eye. Validity bars that are referred to in the text are highlighted in magenta.}
\end{figure}

Figure~\ref{fig:ising6} shows the results for sub-lattices of size $8 \times 8$. At a high temperature of $T = 10^5$, the average DFA exponent across all pairwise comparisons is $0.57$ (see magenta shaded bar). This value is in excess of $0.5$ expected for Gaussian white noise and indicates that even at high temperatures there is order within the rate of change of phase difference between pairs of lattice time series. As the temperature is lowered the DFA exponent of the rate of change of phase difference increases steadily reaching a maximum of $0.65$ at $T=2.55$ (see magenta shaded bar) indicating maximal LRTC just before the critical temperature is reached. 

The change in mean DFA has to be seen within the context of the validity of the DFA fluctuation plots. As the system cools towards the critical point the validity of DFA exponents across all pairwise phase differences drops abruptly. The first temperature value for which $<100\%$ of the DFA plots are valid is $T=2.75$ shown as magenta shaded bar. There is a large fall in DFA fluctuation plot validity as the critical temperature is reached ($56\%$ to $34\%$). This fall in validity reflects the onset of full synchronisation between a number of the time series. At the critical point, $T=Tc$ which occurs between $T=2.25$ and $T=2.3$ (see magenta shaded bars) the validity is $34\%$ of time series pairs with mean DFA exponent of $0.64$. As the Ising model cools below the critical point the DFA validity in general falls and there are no valid DFA fluctuation plots below $T=2.15$. As discussed above this occurs because of the loss of fluctuations in the rate of change of phase difference due to full synchronisation. 

Results obtained for sub-lattice sizes of $32 \times 32$, $16 \times 16$, $12 \times 12$ and $6 \times 6$ were found to be qualitatively consistent with the results shown in Figure~\ref{fig:ising6}  (results not shown).

\subsection{The Kuramoto model}
\begin{figure}
\begin{center}
\includegraphics[scale=0.35]{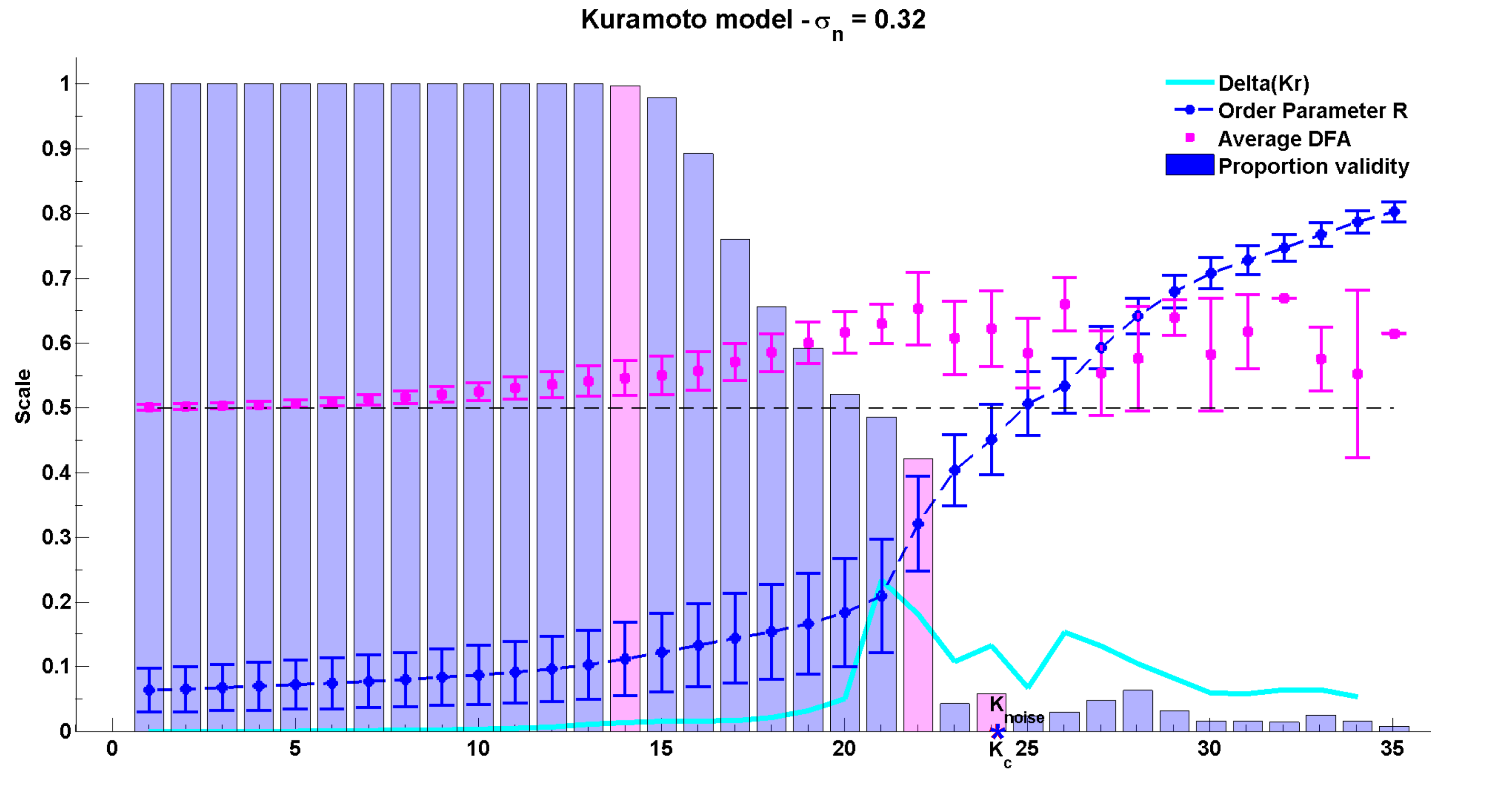}
\end{center}
\textbf{\refstepcounter{figure}\label{fig:kuramoto} Figure \arabic{figure}. Results of the phase synchrony analysis method when applied to the Kuramoto model.}{ There are 200 oscillators, with a mean natural frequency of 22 Hz, and a standard deviation of natural frequencies of 15. The theoretical critical coupling $K_{noise}$ when noise is added is marked with a blue asterisk. The average DFA exponent, order parameter $r$, its difference $\Delta(Kr)$ and the proportion of valid DFA fluctuation plots from the full set of $199000$ pairs are shown. Validity bars that are referred to in the text are highlighted in magenta.}
\end{figure}

The group average results for the Kuramoto model are shown in Figure~\ref{fig:kuramoto}. As can be seen, the peak average DFA exponent occurs on average at $K \approx 22$. The value of the average DFA exponent at this coupling value is $0.65$ with standard deviation $0.06$, consistent with the rate of change of phase difference showing LRTCs. The peak DFA exponent occurs one coupling value later than the peak of the $\Delta(Kr)$ measure, at $K \approx 21$. $\Delta(Kr)$ represents the coupling value at which the order parameter $r$ increases most, and the point of greatest oscillator coupling flux in the system~\citep{bullmore}. The peak coupling value $\Delta(Kr)$ and the maximum DFA values are just less than the theoretical critical coupling of the infinite Kuramoto system with noise $K_c \approx 23.85$. Again, these results must be understood in context of DFA fluctuation plot validity which is $42\%$ of the $199000$ oscillator pairs at $K \approx 22$. Once full synchonization occurs between an individual pair of oscillators, their phase difference takes a constant value. ML-DFA detects the resulting loss of scaling by indicating that the DFA fluctuation plot is no longer linear.

After the peak DFA at $K \approx 22$, further increase in $K$ eventually causes full synchronisation between all individual oscillator pairs. Across the whole system, fewer than $10\%$ of oscillator pairs yield a valid DFA after the critical coupling is exceeded. When all oscillator pairs are synchronised with each other, the order parameter of the system approaches its maximum level of $1$ but the DFA fluctuation measure of rate of change of phase difference is no longer valid. 

Analysis of the Kuramoto model with noise suggests that LRTCs in the rate of change of phase difference between oscillator pairs occur when the system is in a state of maximal flux just prior to the onset of full synchronisation.

\subsubsection*{Individual oscillators pairs.}
\begin{figure}
\begin{center}
\includegraphics[scale=0.45]{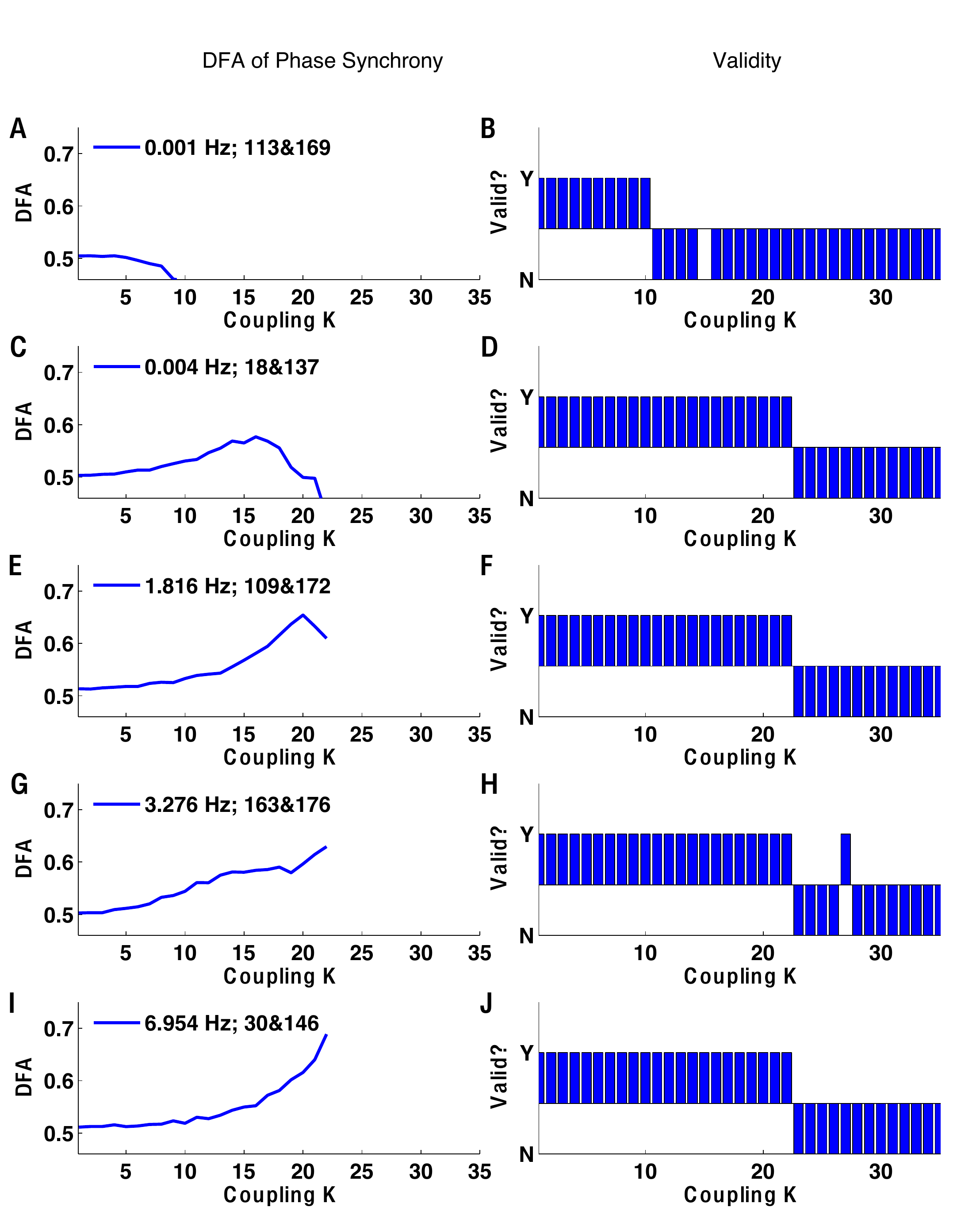}
\end{center}
\textbf{\refstepcounter{figure}\label{fig:expev} Figure \arabic{figure}. Representative relationship of DFA exponents to the coupling parameter $K$ for selected oscillator pairs in the Kuramoto system.}{ Panels A,C,E,G,I and K show the value of valid DFA exponents, while panels B,D,F,H,J and L indicate whether the exponent is rejected as invalid by the ML-DFA technique (N) or not (Y). The oscillator numbers and the differences between their natural frequencies are recorded in the legend of panels A,C,E,G,I and K. The first number is the difference in natural frequency (in Hz), and the subsequent pair of numbers identifies which oscillators are being analysed.} 
\end{figure}

Further insights into the rate of change of phase difference fluctuation behaviour can be obtained from DFA of individual oscillator pairs. Analysis of a set of 5 oscillator pairs is shown in Figure~\ref{fig:expev}. The top panel shows the change in DFA exponent with coupling $K$ for a pair whose initial frequencies are very close ($0.001$Hz apart). The bottom panel shows the changes in DFA exponent for an oscillator pair with initial frequencies that differ by $\approx 7.0$Hz. The middle panels show oscillator pairs with varying amounts of initial frequency difference (increasing top to bottom). Non valid DFA exponents are not plotted in the left hand panel but the right hand panels indicate for each given pair linear DFA validity 'yes' or 'no' for a given value of $K$. At low coupling $K$, the oscillators do not interact with each other and each evolves at its own natural frequency. The order in the system is low and the DFA exponent $\approx 0.5$ reflects the additive noise which dominates the fluctuations in the rate of change of phase difference. A DFA value of $\approx 0.5$ is also evident in the average DFA (Figure~\ref{fig:kuramoto}). There is almost $100\%$ validity across all pairs because white noise time series are scale-free and therefore the DFA fluctuation plot obtained from analysing them is expected to be linear (Figure \ref{fig:kuramoto}).

As the coupling parameter $K$ is increased, the DFA exponents of each of the oscillator pairs rise until a peak is reached. The value of $K$ at which a maximal valid exponent is retrieved for these peaks is related to the difference in natural frequencies of the two oscillators as well as their interactions with the noise and the mean field. Oscillator pairs which start further apart in frequency terms develop full synchonization later than those whose initial frequencies are close together. As $K$ increases the DFA exponent of the rate of change of phase difference increases. The pairs with the strongest LRTCs on the basis of the highest DFA exponent value prior to onset of full synchronisation are those with the greatest inital frequency difference. Increasing temporal order of the rate of change of phase difference prior to full synchonization of these pairs may indicate a state of pre-synchronisation in these pairs.

\subsection{The Cabral model}
For the Cabral model we present results regarding both the global behaviour of the system through average DFA exponents across all possible pairs of oscillators (Figure~\ref{fig:cabral}) and the behaviour of the system at cluster level through average DFA exponents of intra-cluster pairs of oscillators (Figure~\ref{fig:intracluster}).

\subsubsection{Global behaviour.}
\begin{figure}
\begin{center}
\includegraphics[width=15cm]{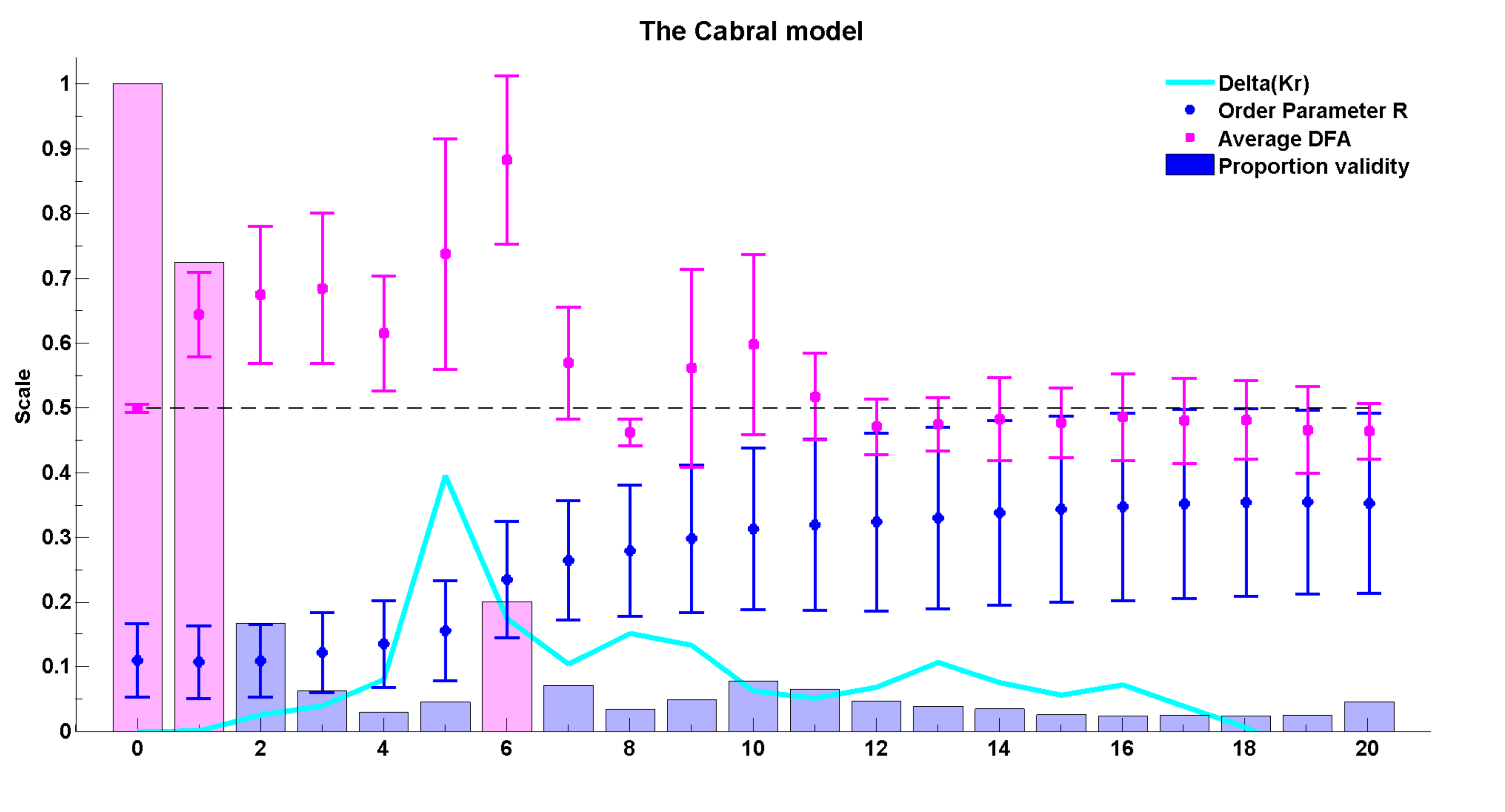}
\end{center}
 \textbf{\refstepcounter{figure}\label{fig:cabral} Figure \arabic{figure}. The average DFA exponents of phase synchrony as a function of the coupling parameter, $K$, in the extended Kuramoto model \citep{cabral}.}{ The model includes $66$ oscillators at normally distributed natural frequencies with mean 60 Hz and standard deviation $\sigma_i = 1.25$. The connectivity and time delay matrices are set from empirical values. The average of the valid DFA exponents is shown in magenta and the proportion of valid exponents, as calculated by ML-DFA, are indicated by bars. The Kuramoto model order parameter $r$ is in blue, and the quantity $\Delta(Kr)$ is in cyan. The peak $\Delta(Kr)$ has been used as an indicator of the effective critical coupling. A horizontal line at DFA exponent 0.5 is plotted to guide the eye. The proportion of valid DFA bars for $K=0$, $K=1$ and $K=6$ have been shaded in magenta.}
\end{figure}

The model introduced by~\citet{cabral} is affected by rich interplay between the connectivity and distance matrices as well as the noise and natural frequency elements of the system. The average valid DFA exponents for all oscillator pairings (n=$2145$) are shown in Figure~\ref{fig:cabral} as the coupling in the system is increased. These average exponent values indicate the presence of LRTCs in the rate of change of phase difference. The peak values of mean DFA exponent correspond to peaks in the change in order paramenter ($\Delta(Kr)$) derived for the classical Kuramoto model and the Kuramoto model with noise, see~\citet{bullmore} and Figure~\ref{fig:kuramoto}. Such peaks occur when the system undergoes the greatest change in synchronisation. The peak in $\Delta(Kr)$ corresponds closely to the coupling value that shows maximum mean DFA exponent ($K=5$ and $6$, respectively -- see Figure~\ref{fig:cabral}).

The number of pairings that yield valid DFA exponents in the rate of change of their phase difference is equal to $100\%$ when there is no coupling in the system (magenta shaded bar at $K=0$), but it falls as coupling is introduced (magenta shaded bar at $K=1$). At the coupling value of the DFA peak, $K=6$, validity is at $20\%$, which is higher than the neighbouring coupling values (magenta shaded bar at $K=6$). 

\subsubsection{Cluster behaviour.}

\begin{figure}
\begin{center}
\includegraphics[width=15cm]{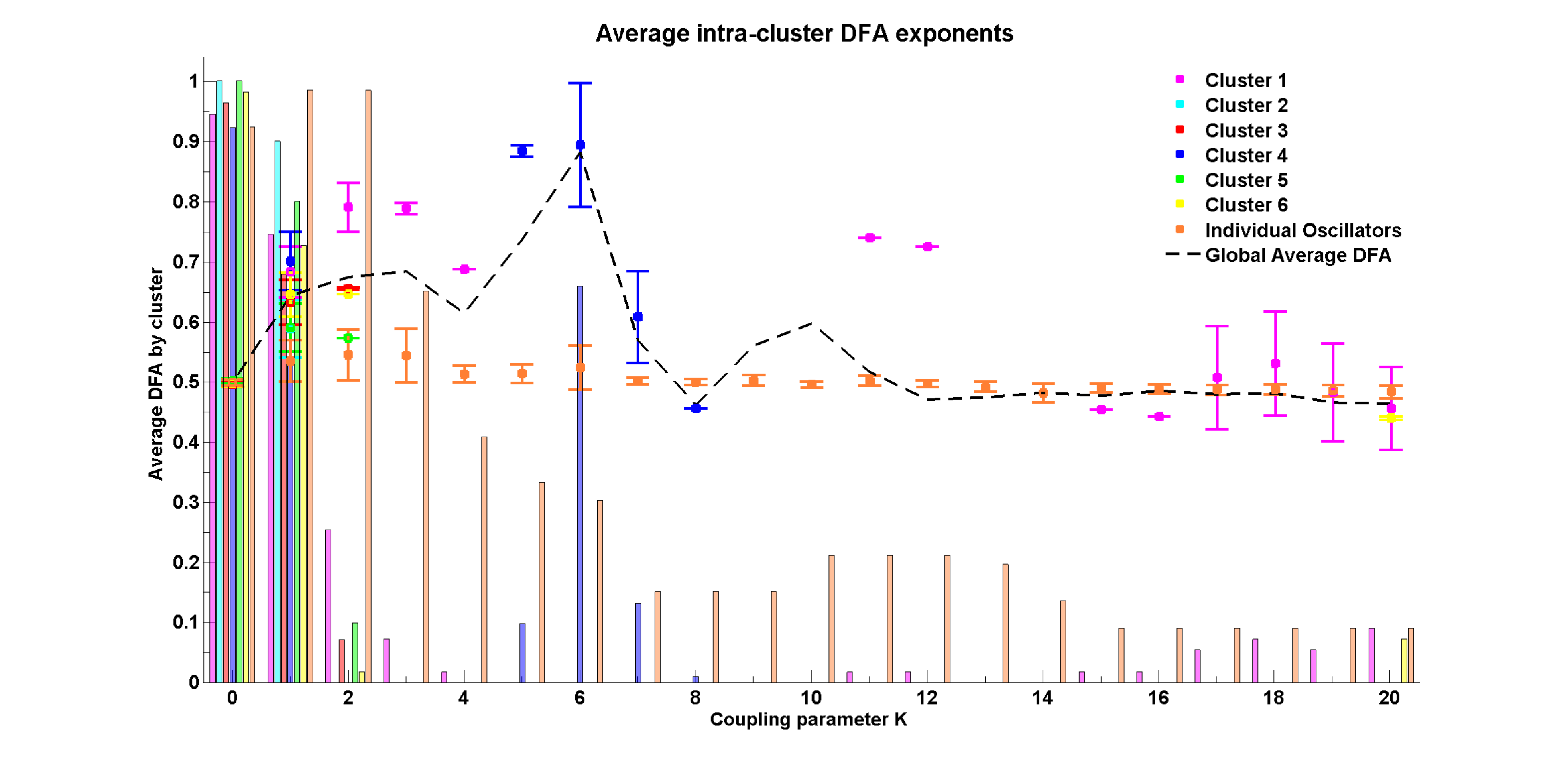}
\end{center}
 \textbf{\refstepcounter{figure}\label{fig:intracluster} Figure \arabic{figure}. Average DFA exponent for intra-cluster pairwise phase differences with increasing coupling parameter $K$.}{ Where no DFA value appears for a particular cluster, this indicates that there are no valid DFA exponents for the pairwise phase difference within that cluster. The final cluster, which is labelled individual oscillators, consists of a set of nodes that do not fit into any of the clusters as determined by the connectivity and distance matrices but are grouped together to demonstrate their relationship with each other. }
\end{figure}

At coupling value $K=6$, the value at which the global behaviour shows peak DFA value, the intra-cluster results indicate that only cluster 4, consisting of oscillators 27-40, shows valid non-trivial DFA exponents. These exponents are consistent with the presence of LRTCs. This suggests that cluster 4 acts as an organising force in the system when the system is in its greatest state of flux, as demonstrated by a large increase in the order parameter. This cluster corresponds to the most connected brain regions listed in Table~\ref{tab:clusters} and Table~\ref{tab:brains} in the Appendix. 

\begin{figure}
\begin{center}
\includegraphics[width=15cm]{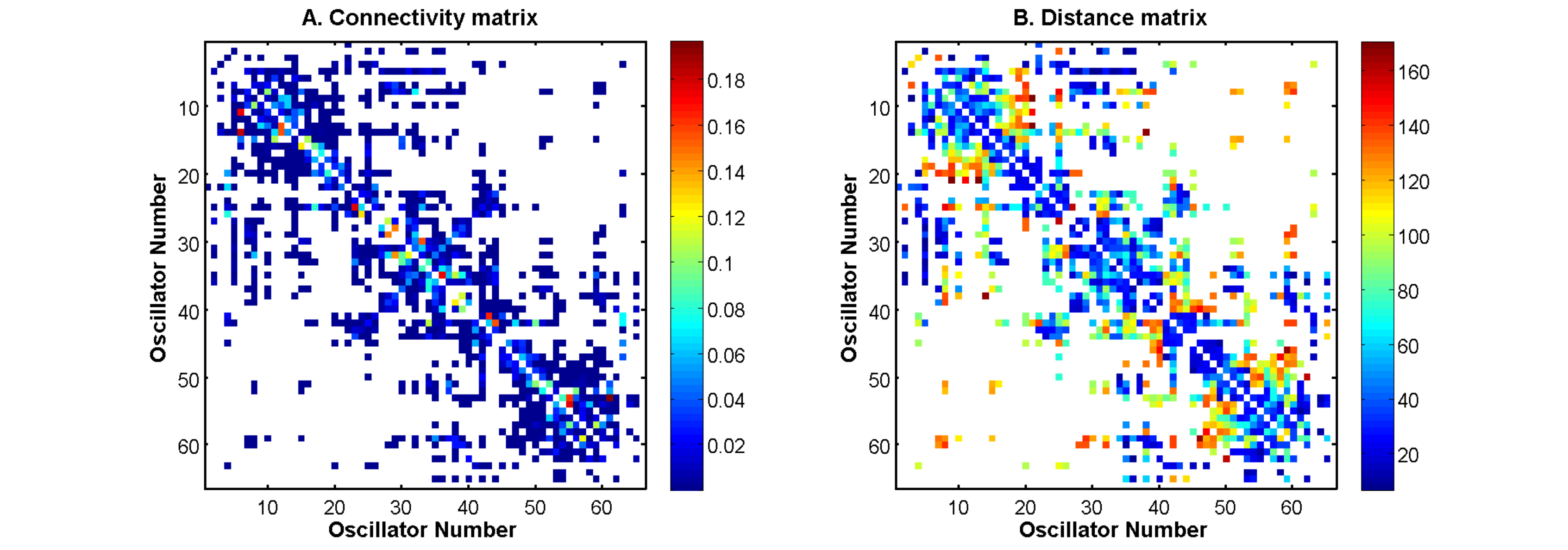}
\end{center}
 \textbf{\refstepcounter{figure}\label{fig:DFAConnDist} Figure \arabic{figure}. Connectivity and distance matrices for the Cabral model.}{ Each oscillator number represents a brain region, which is defined in Table~\ref{tab:brains} in the Appendix. An empty (white) element means that the two regions are not connected. Regions are not connected to themselves so that the diagonals are white. Panel A shows the pairwise connection matrix $C$ between the 66 oscillators. Panel B shows the matrix of pairwise distances $L$ between the brain regions that are represented by the 66 oscillators. Matrix $L$ is symmetric, however, matrix $C$ is not because the connection weights are normalised by row. The values associated with the colours of the plots are defined by the colour bars. Red colours in panel A represent higher weights. Red colours in panel B represent longer distance connections.}
\end{figure}

\begin{figure}
\begin{center}
\includegraphics[scale=0.40]{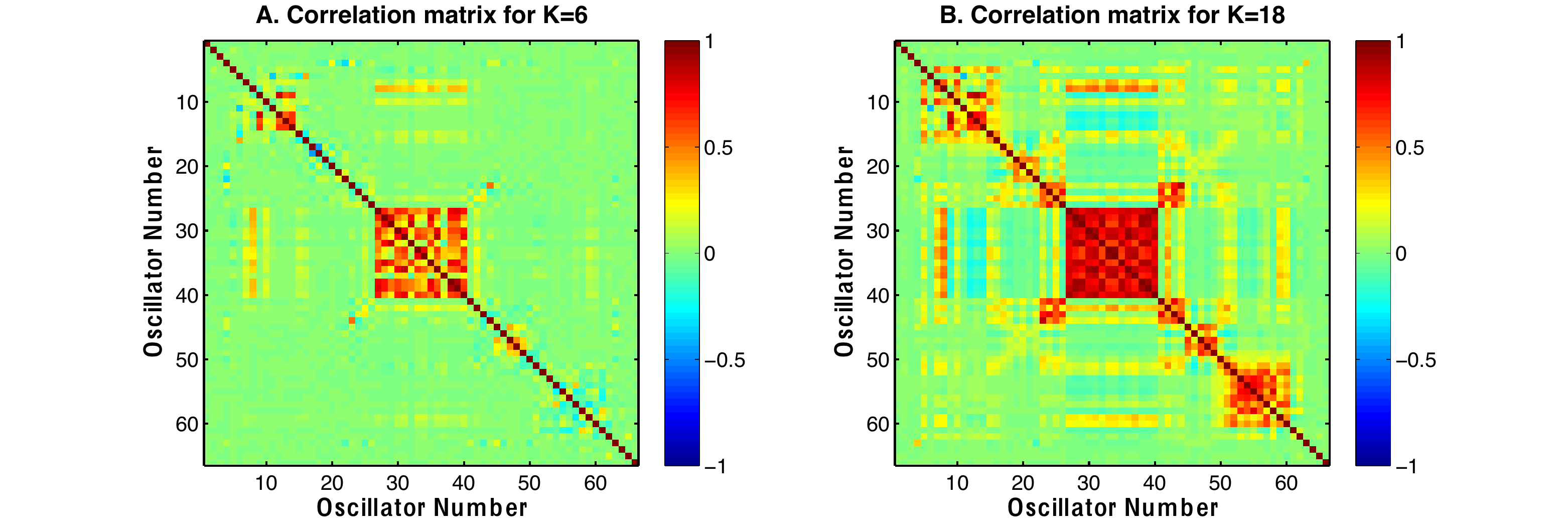}
\end{center}
 \textbf{\refstepcounter{figure}\label{fig:fig13} Figure \arabic{figure}. Correlation matrices for all pairs of time series generated by the Cabral model for two coupling values $K$.}{ A. $K=6$ and B. $K=18$, which corresponds to the oscillator correlation matrix in~\citet{cabral}. The plots show the value of the Pearson correlation coefficient between all pairwise combinations of the $66$ oscillators used in the model.}
\end{figure}

The connectivity and distance matrices for the Cabral model are shown in Figure~\ref{fig:DFAConnDist}. The linear coupling between oscillators for two values of $K$ is shown in Figure~\ref{fig:fig13}. The central cluster of oscillators with high levels of synchronization is evident from the two correlation matrices.   At $K=6$ (panel A), i.e., the value at which LRTCs are detected in the rate of change of phase difference, the central oscillator cluster shows evidence of synchronization but with Pearson correlation values of $<1.0$.  As $K$ increases to 18, the value identified by~\citet{cabral} as best approximating human brain resting state BOLD fluctuations, it can be seen from Figures~\ref{fig:cabral} and~\ref{fig:intracluster} that the proportion of oscillator pairs with valid DFA fluctuation plots is low (approximately $5\%$). Those oscillator pairs that remain and show persistently valid DFA fluctuation plots are predominantly individual oscillators with low average weight per node (0.03) and low average degree distribution (8.59).  Their associated DFA exponent is on average 0.5 (see Figure~\ref{fig:intracluster}).  At $K=18$, the Cabral model shows strong cluster synchronisation. In particular, the central cluster 4 (oscillators $27-40$) which contains homologous elements connected across the corpus callosum shows Pearson correlation values close to 1.0 indicative of full synchrony (Figure~\ref{fig:fig13}B). Therefore the results we obtained from the Kuramoto model with noise and those derived from the Cabral model are similar.  Both show valid DFA fluctuation plots with LRTCs of the rate of change of phase difference at a coupling value where $\Delta (Kr)$ is increasing and loss of validity as full synchronisation takes over.  As discussed earlier, 'criticality' is not defined for the Cabral model but with increasing $K$ there is clearly a change in the system's order which is detected through our method.

Figure~\ref{fig:Kis18} shows the DFA exponents of the rate of change of phase difference between individual pairs of oscillators in the form of a symmetric lattice of size $66 \times 66$, where each element in the lattice represents a brain region as detailed in Table \ref{tab:brains} of the Appendix. Panel A of this figure shows the importance of the central cluster in generating LRTCs of phase synchronisation. Importantly it shows this cluster's influence over many of the other oscillators in the Cabral model. Cluster group 4 has the greatest sum of weights per oscillator and the greatest number of connections per oscillator (see Table~\ref{tab:clusters}). The correlation between the number of connections of a given oscillator and the average DFA exponent of its rate of change of phase difference with all other oscillators is $0.359$, suggesting a relationship between oscillators with large connectivity and those with large DFA exponents in their pairwise phase difference. 

\subsubsection{Comparison of the three connectivity structures.}
\begin{figure}
\begin{center}
\includegraphics[width=15cm]{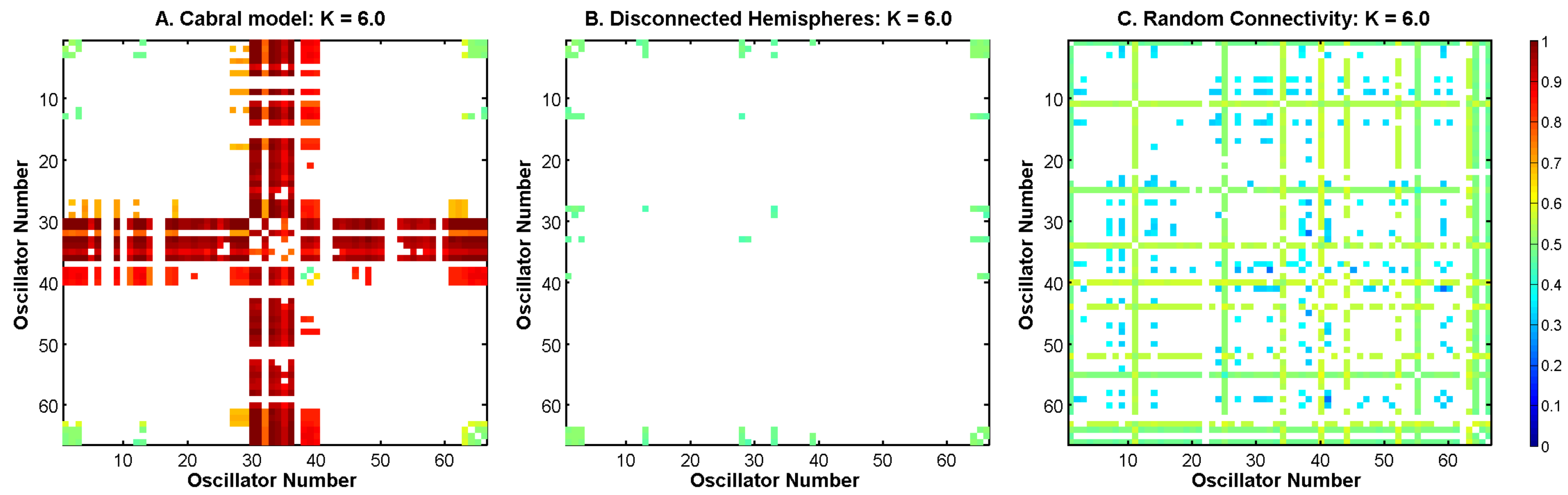}
\end{center}
 \textbf{\refstepcounter{figure}\label{fig:Kis18} Figure \arabic{figure}. DFA exponent of the rate of change of phase difference between all pairs of oscillators in the Cabral model at coupling $K=6$ in three scenarios.}{ A: For the empirically observed connectivity matrix of the Cabral model. B: For a connectivity matrix representing disconnected hemispheres. C: for random connectivity. Empty (white) elements denote pairs for which no valid DFA exponent was found.}
\end{figure}

In the Cabral model, the $\Delta(Kr)$ measure has its peak at coupling value $K=6$. Here, we compare the effects of the three connectivity matrices introduced in Section~\ref{sec:cabraldisrupted} on the DFA exponents of the pairwise phase difference between oscillators at this coupling value in Figure~\ref{fig:Kis18}. 

The empirical connectivity matrix showed large DFA exponents indicating the presence of LRTCs at this coupling value for a small number of hub oscillators belonging to cluster 4 (see above). These oscillators have a high number of connections and large weights associated with these connections (see Table \ref{tab:clusters}). When the two hemispheres are disconnected, we see no LRTCs in the DFA exponents of the phase difference at this coupling value. When the distance matrix is preserved, but the connectivity and associated weights are assigned at random, LRTCs are still present in the DFA exponent of the phase differences between oscillators, but a lower value of DFA exponent is obtained. There is no apparent cluster formation when connectivity is random.

\subsubsection{Neurophysiological data.}

\begin{figure}
\begin{center}
\includegraphics[width=15cm]{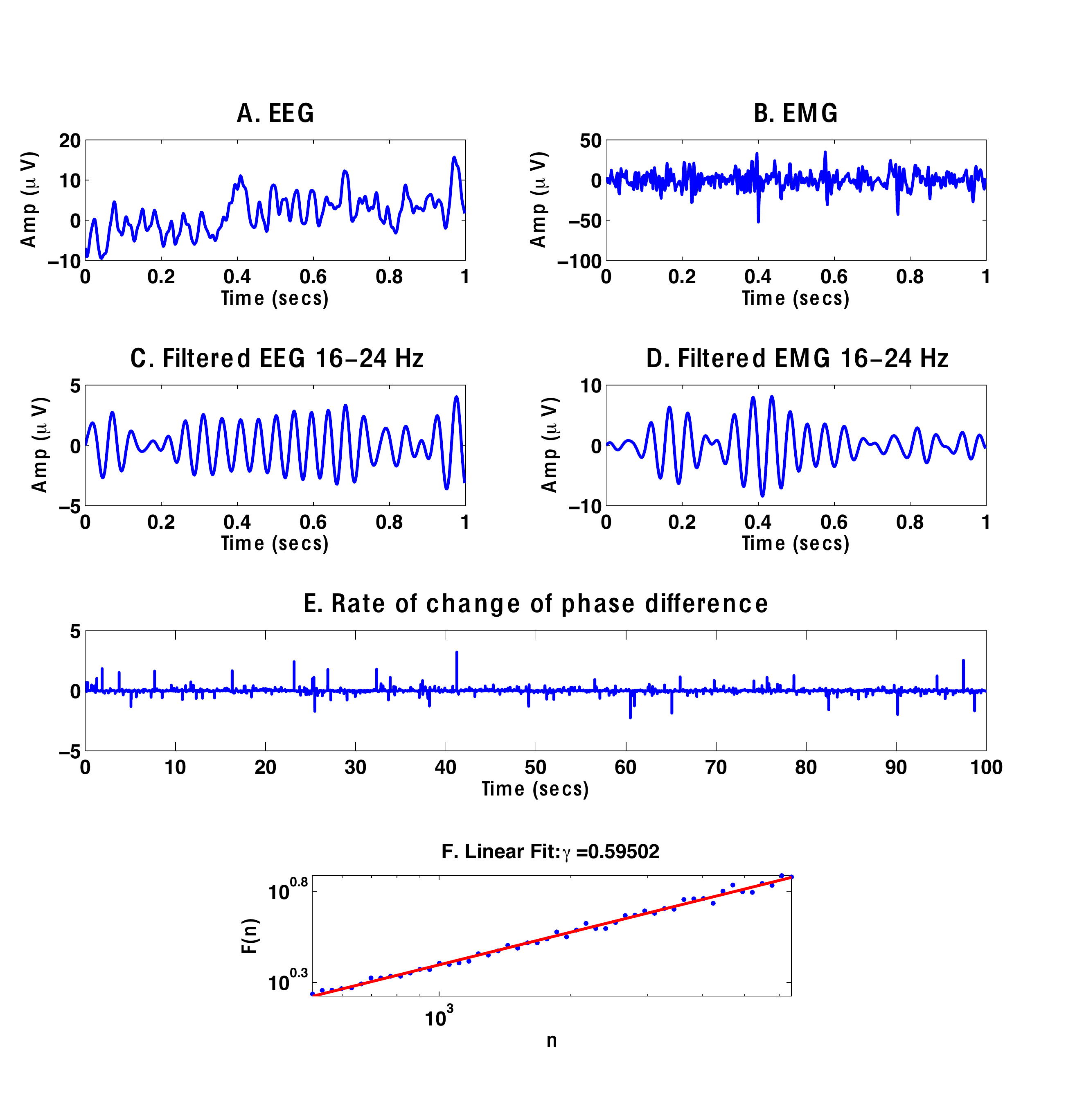}
\end{center}
 \textbf{\refstepcounter{figure}\label{fig:eegemg} Figure \arabic{figure}. Illustration of the method with simultaneous EEG/EMG data.}{ A and B: One second of simultaneously recorded EEG and EMG, respectively. C and D: The signals after bandpass filtering in the $\beta$ range $16-24$Hz. E: Rate of change of the phase difference between the two bandpass filtered signals for 100 seconds. F: DFA fluctuation plot for the rate of change of phase difference time series in panel E. The plot was determined to be valid by ML-DFA with a DFA exponent of $\approx 0.60$, indicating the presence of LRTCs. }
\end{figure}

Figure~\ref{fig:eegemg} illustrates the application of our phase synchrony analysis technique to the human neurophysiological data described in Section~\ref{sec:physiodata}. In this example, a valid DFA exponent of $\approx 0.6$ was obtained for the rate of change of phase difference between the simultaneously recorded EEG and EMG data during a steady muscle contraction, indicative of the presence of LRTCs. Analysis of amplifier noise and artificially generated noise time series using processing steps identical to those for the EEG and EMG data (signal processing pathway shown in Figure~\ref{fig:method}) resulted in a valid DFA fluctuation plot but with exponent of 0.48 consistent with uncorrelated noise. 

\section{Discussion}
The aim of this paper is to introduce a new methodology for eliciting a marker of criticality in neuronal synchronisation. This methodology relies on the rate of change of the phase difference between two signals as a (time-varying) measure of phase synchronisation. The presence of long-range temporal correlations in this quantity is proposed as marker of criticality and is assessed using detrended fluctuation analysis (DFA) in combination with the recently proposed ML-DFA, a heuristic technique for validating the output of DFA. With these methods, we can first determine the presence or absence of power law scaling using ML-DFA and secondly the presence or absence of long-range temporal correlations (LRTCs) in the phase synchronisation of two time series based on the value of the DFA exponent. If the method returns an exponent of $\approx 0.5$, this indicates a phase relationship similar to white Gaussian noise, however, if the DFA exponent is greater than 0.5, this indicates the presence of LRTCs.  Importantly, we can attribute significance to the loss of power law scaling within the fluctuation plot and draw conclusions based on an exponent value only when the exponent has been recovered from plots that are judged to be valid by ML-DFA.

\subsection*{Surrogate Data}
It was found that the phase synchrony analysis method recovers a known DFA exponent value in the rate of change of phase difference between two signals of surrogate data with a high degree of accuracy (r=0.998). When the structure of phase synchronisation was perturbed with an additive noise source, it was found that a percentage difference between the true and recovered DFA exponent of above approximately $5\%$ noise caused DFA exponents to be judged as invalid by ML-DFA.  When the surrogate data was characterised by a DFA exponent close to 1, the recovery of this exponent using DFA was more resistant to noise when compared to surrogate data with a lower DFA exponent of 0.6 (Figure~\ref{fig:surrogate5}). In these simulations we used additive noise which was included at the amplitude stage of the surrogate time series prior to extraction of the phase using the Hilbert transform. 

\subsection*{The Ising model}
We had initially expected to see LRTCs in the Ising model only in the vicinity of the critical parameter, and a DFA exponent of 0.5 when the energy in the system was large (disordered phase). However, in applying our method to the Ising model, both of these hypotheses were not fully realised. It was found that when the temperature was increased to a very high level of $T=10^5$, the DFA exponent of the rate of change of phase difference did not fall to 0.5, but remained at $\approx 0.57$. This did not change when the temperature was set to an even higher value of $T=10^{12}$. This was not a finite size effect of the system, as the result held when larger lattice sizes (up to $1000 \times 1000$) were used (results not shown). We noted that when pure phase was analysed, i.e., an uncoupled system of Kuramoto oscillators, DFA exponents of 0.5 were obtained as expected, and therefore, we cannot exclude the possibility that the Hilbert transform induced artefacts may inject some order into the resulting phase time series. However, within the Ising system, the expectation of a DFA exponent of 0.5 at high $T$ is based only on our intuition concerning the operation of the system. As all elements in the Ising lattice interact with their neighbours it is possible that some temporal correlation in the rate of change of phase difference may persist regardless of temperature value, and this may be the cause of a DFA exponent above 0.5. 
 
Importantly, we found that the DFA exponent was indicative of LRTCs at critical temperature but was maximal at $T=2.55$, just in excess of the critical temperature.  As can be seen in Figure~\ref{fig:ising6}, the consistent change in the DFA value and the change in power law scaling behaviour indicates that the phase synchrony analysis method is capturing an important behaviour of the system close to its critical regime.  However, it is important to realise that unless an experimental neuroscientific paradigm can be discovered that produces similar consistent changes in this measure, neurophysiological data will have to be intepreted with caution, i.e., we may be able to state that for a given pair of neural oscillation time series there exists power law scaling with a DFA exponent indicative of LRTCs in the rate of change of their phase difference but we may not know whether for this neural state there may exist other higher (or lower) exponent values. In other words, the technique may provide evidence that the system is ordered in ways that are similar to systems nearing their critical regime but whether the technique will pinpoint the most critical regime in a neural system is open to question. We will consider this further in our discussion of the results of analysing a Kuramoto system with noise.

Interestingly, the evolution of the DFA exponent with the temperature parameter shares a key characteristic with that of a recently published measure of information flow in the same model~\citep{barnett}, specifically, an asymmetry around the critical point, with a sharp rise in the metric as temperature is increased towards the critical $T=Tc$ and a gradual descent as the temperature rises significantly. It would be of interest to further assess the extent to which the proposed method captures information flow in the system, e.g., through a comparison of both methods when applied to the Kuramoto model. 

\subsection*{The Kuramoto model}
In the Kuramoto model, the critical transition is characterised in terms of a global order parameter which reflects the overall organisation of the system.  However, through our phase synchrony analysis method we are able to make observations at a pair-wise level of Kuramoto oscillators always bearing in mind that even at the pair-wise level the result is influenced by the oscillators' interactions with all other oscillators in the model. As individual Kuramoto oscillator pairs become fully synchronised, their rate of change of phase difference no longer contains moment-to-moment fluctuations and thus power law scaling in the DFA measure is lost. This is an important consideration because it emphasises the difference between our method and more standard measures of neural synchrony.  Methods for detecting neural synchrony rely on phase consistency to allow averaging out of fluctuations so that a measure of coupling (e.g., coherence and phase coherence) is obtained.  In contrast, the method introduced in this paper is dependent on the fluctuations of the two phase signals and their interaction.  Therefore our method detects  'order' across time in the rate of change of phase difference rather than phase consistency between two processes. 

The phenomenon of loss of fluctuations at the onset of full synchronisation is well illustrated both for the global Kuramoto model and for individual oscillator pairs extracted from the Kuramoto model.  In the global analysis the peak in the DFA exponent occurs close to the observed peak of $\Delta(Kr)$ and at values of $K$ just below theoretical critical coupling value. At these values of $K$, a power law scaling exists for the rate of change of phase difference, and the DFA exponent of oscillator pairs with different initial frequencies indicates the presence of LRTCs. At the onset of full synchronisation the number of oscillator pairs for which DFA is valid drops yet those whose phase differences still possess fluctuations continue to show LRTCs.  Once the critical regime has been fully crossed and the order parameter $r$ approaches 1, the DFA of the rate of change of phase difference is no longer valid for any oscillator pair.
  
The LRTC behaviour is also clearly explained as the coupling value $K$ decreases towards zero.   As can be seen in Figure~\ref{fig:kuramoto}, the DFA exponent of the pairwise rates of change of phase difference decreases towards 0.5 and yet scaling remains valid. These changes in DFA exponent are evident both on the global level in the average DFA and for individual oscillator pairs.  At $K=0$ the phases are independent from one another yet contain noise; thus the rate of change of phase difference time series contains innovations that are random across time with a DFA which is valid and returns the expected exponent of 0.5.
 
\subsection*{Order within the Ising and Kuramoto models}
In these models, temperature $T$ (Ising) and coupling $K$ (Kuramoto) play a similar role in controlling the \emph{order} within the two systems, and the DFA validity and exponent results obtained from analysis of rate of change of phase difference in both of these models mirror each other. In the Kuramoto model, there is a transition from an uncoupled to a synchronised state with increasing $K$. Similarly in the Ising model, there is a transition from a very ordered to a disordered system with increasing $T$. In the human brain, we are not able to characterise the system by incrementally tuning a parameter and observing the result, and we are only privy to snapshots of the working system. However, we can begin to understand the behaviour of the brain within this range of behaviours by comparing the DFA of the rate of change of phase difference of pairs of neurophysiological signals to the outcomes of these models of criticality.

\subsection*{The Cabral model}
We found that LRTCs exist in the rate of change of phase difference between oscillator pairs at parameter values close to those at which the change in order, $\Delta (Kr)$, increases sharply. Extrapolating from the Kuramoto model with noise, we suggest that there are important changes in the order of the phase synchronisation of interacting oscillators in the Cabral model that involve the presence of LRTCs when the order in that system is at or close to a point of maximal change. 

It is important to note that the value of $r$ in the Cabral model does not reach a level of 1 in the range of coupling values $0-20$. It approaches a level of $\approx 0.4$ as $K$ approaches 20 with maximal rate of change at $K \approx 6$. Further analysis of the Cabral model indicates that $r$ will gradually reach a value closer to 1 as $K$ increases above a value of $60$, as seen in Figure 4 of ~\citet{cabral}. Cabral focussed her attention on $K=18$ at which point the model, when fed through the Balloon-Windkessel hemodynamic model, produced an output that closely matched the spatio-temporal correlations seen in the BOLD signals of the resting state fMRI. We find that at this value, there are no LRTCs detectable in the rate of change of phase difference measure. 

\subsection*{The role of connectivity in the Cabral model}
Although most of results were obtained at $K=6$, selected because it is the peak of $\Delta (Kr)$, it is important to note that LRTCs exist for a broader range of coupling values $K$. This finding agrees with a recent study by~\citet{moretti} in which the authors demonstrated that a network with complex connectivity, such as that of the Cabral model and, indeed, that of the brain, causes the critical point to becomes a broader critical 'region'.

Our examination of oscillator pairs belonging to a single cluster, as defined in~\citet{cabral}, indicates that the emergence of LRTCs is determined primarily by oscillators belonging to cluster 4 which has a large number of connections and a large sum of connection weights. This cluster is located centrally, and it contains four brain regions of particular importance to the resting state network~\citep{fransson,richclub}. These are oscillators 33 and 34, which correspond to the left and right posterior cingulate cortices, and oscillators 32 and 35 which represent the left and right precuneus.  These central brain regions are known to be important with a higher metabolic activity than other regions during the resting state. 

Importantly, we find that LRTC behaviour of this cluster, and its relationship to the other clusters in the network, is dependent on trans-callosal left-right connectivity. Indeed, disruption of the left-right trans-callosal connections resulted in a loss of LRTCs in the rate of change of phase difference between time series extracted from the central cluster 4 and the other oscillators in the Cabral network. Intuitively, those oscillators that are connected to many other oscillators in the network will also influence the phases of a large number of other oscillators. When these oscillators try to synchronise, we suggest that those that are well connected will be subjected to conflicting phase inputs from their neighbours and thus increased variation in their phase fluctuations, yielding a larger DFA exponent. These variations in fluctuation will in turn feed into the neighbouring oscillators and cause them to also have large variations in fluctuation as they attempt to synchronise with their well-connected neighbour. On the other hand, an oscillator that is poorly connected or connected to just one other oscillator may have a more straightforward task of synchronising with just this (albeit changing) oscillator speed.

The LRTCs in the rate of change of phase difference were also disrupted by randomisation of connectivity, albeit less severely than when the trans-callosal connections were severed. When a random connectivity is assigned, no clusters exist and DFA exponents are significantly reduced.

The results obtained from the phase synchrony analysis method here may pave the way for potential future use of the Cabral model in investigating specific pathological modifications of connectivity and their effects on the time-varying synchronisation patterns between different brain regions. The method has the potential to be used to trace some types of pathological synchronisation such as may arise in epileptic or Parkinsonian conditions to any roots that they may have either in the connectivity, clustering or noise input elements of the Cabral model and therefore potentially also of the nervous system. 

\subsection*{Neurophysiological data}
In order to show proof of principle, we have presented an example of our method's application to neurophysiological data, in this case EEG and EMG simultaneously recorded during voluntary muscle contraction. It was through this experimental paradigm that corticomuscular coherence (CMC) in the $16-32$Hz ($\beta$) frequency range was first discovered by~\citet{conway95,halliday98} and shown to be the $\beta$ frequency common drive to human motoneurons first described by~\citet{farmer93}. 
These preliminary results indicate power law scaling in the DFA plot with a DFA exponent of $\approx 0.6$. 

It has been recognised through application of time-varying coherence measures that CMC coherence fluctuates even when a subject attempts to maintain the same motor output~\citep{muthu11}. As discussed earlier, the techniques introduced here allow us to focus on the fluctuations within the phase coupling rather than on the averaged measure of coupling.  These preliminary results indicate that the fluctuations in the rate of change of phase difference between simultaneously recorded EEG and EMG show power law scaling and LRTCs within the $\beta$ frequency range. 
We suggest that the analysis of instantaneous phase diffence of neurophysiological data using the methods described in this paper will allow researchers to investigate the coupling between signals in a way that will allow a new appreciation of the relationship between neural synchrony and other oscillator systems approaching their critical regime.

\subsection*{LRTCs in rate of change of phase difference and the brain}
LRTCs have been associated with model dynamical systems that show efficiency in learning, memory formation, rapid information transfer and network organisation. The broad dynamical range of which LRTCs are a marker acts to support these functions~\citep{linkenkaer01,chialvo,sornette,timme,stam,werner2010,linkenkaer04,meisel,shew}. It has been argued by a number of researchers that these properties if present would be of major benefit to the functions that human brain dynamics needs to support and there is now a literature that connects the theory of critical systems with properties of human brain dynamics~\citep{bullmore,chialvo,shew,linkenkaer01,beggsandplenz}. 

In this paper, we focus on LRTCs, and because of the importance in neuroscience of brain oscillations and the concept of communication through coherence, we make the link between LRTCs and phase synchrony. We note that in the model systems that we have explored the highest valid DFA exponents were recovered when the systems were close to their critical point but in a slightly more disordered state than at exact criticality. We explained this on the basis of full synchronisation within our model systems being a point at which the rate of change of phase difference is lost (observed in Ising at $T<T_c$ and in Kuramoto for increasing $K$). 

In neurophysiological systems, it is important to appreciate that full synchronisation of neural oscillators is a pathological state (e.g., observed in the EEG and MEG of epileptic seizures and in EMGs showing pathological tremor). The healthy resting brain state therefore is characterised by weak and variable neural synchrony which would be expected to show fluctuations (temporal innovations) in a measure of the change in phase synchrony, i.e., the rate of change of phase difference. From the perspective of brain dynamics (and muscle activation dynamics) the most important constraints are to avoid pathological synchronisation whilst at the same time maintaining the potential for useful synchronisation. We suggest therefore that in the healthy state the instantaneous phase difference between neural oscillators will show power law fluctuation plots with a DFA exponent that is either 0.5 or that will show LRTCs. If LRTCs are found in the resting state then they may represent an optimum state of readiness to which the system can readily return if increased synchronisation occurs as a result of sensory stimulation, motor task or cognitive action. Such temporary changes in synchronisation may occur in order to support communication through coherence. The resting state, however, is characterised by fluctuations of phase synchrony that have LRTCs and represent the behaviour of weakly coupled oscillators whose synchrony can be modulated. The hypothesis that the LRTCs of rate of change of phase difference of brain oscillations may be altered through task is an experimentally tractable question. 

To conclude the evidence for the brain as a critical system continues to accrue. There is an important need to link the criticality paradigm with the paradigm that attaches functional significance to neural synchrony. The methodology presented in this paper takes us some way towards this synthesis. 

\section*{Acknowledgement}
Maria Botcharova thanks the Centre for Mathematics and Physics in the Life Sciences and Experimental Biology (CoMPLEX), University College London for their funding and continuing support. Simon F. Farmer was supported by University College London Hospitals Biomedical Research Centre (BRC). 

\appendix

\begin{table}[!ht]
\caption{{\bf A list of the 66 brain regions which are represented by 66 oscillators in the Cabral model.} The abbreviations, full names and oscillator numbers corresponding to the left and the right hemispheres are given for each brain region. }  
\centering 
\begin{tabular}{|l|l|l|l|l|l|l|l|l|l|l|l|l|l|l|l|l|l|l|l|l||l|l|l|l|l|l|l|l|l|l|l|l|l|l|l|l|l|l|l|} 
\hline 
  &   &  \multicolumn{2}{c|}{Oscillator number} \\
\hline
Abbreviation&Region&Right&Left\\
[0.5ex]
\hline 
ENT & Entorhinal cortex &1&66\\
PARH & Parahippocampal cortex&2&65\\
TP& Temporal pole&3&64\\
FP& Frontal pole&4&63\\
FUS& Fusiform gyrus&5&62\\
TT& Transverse temporal cortex&6&61\\
LOCC& Lateral occipital cortex&7&60\\
SP& Superior parietal cortex&8&59\\
IT& Inferior temporal cortex&9&58\\
IP& Inferior parietal cortex&10&57\\
SMAR& Supramarginal gyrus&11&56\\
BSTS& Bank of the superior temporal sulcus&12&55\\
MT& Middle temporal cortex&13&54\\
ST& Superior temporal cortex&14&53\\
PSTC& Postcentral gyrus&15&52\\
PREC& precental gyrus&16&51\\
CMF& Caudal middle frontal cortex&17&50\\
POPE& Pars opercularis&18&49\\
PTRI& Pars triangularis&19&48\\
RMF& Rostral middle frontal cortex&20&47\\
PORB& Pars orbitalis&21&46\\
LOF& Lateral orbitofrontal cortex&22&45\\
CAC& Caudal anterior frontal cortex&23&44\\
RAC& Rostral anterior cingulate cortex &24&43\\
SF& Superior frontal cortex&25&42\\
MOF&Medial orbitofrontal cortex&26&41\\
LING& Lingual gyrus&27&40\\
PCAL& Pericalcarine cortex&28&39\\
CUN& Cuneus&29&38\\
PARC& Paracentral lobule&30&37\\
ISTC & Isthmus of the cingulate cortex&31&36\\
PCUN& Precuneus&32&35\\
PC& Posterior cingulate cortex&33&34\\
\hline 
\hline 
\end{tabular}
\begin{flushleft} The labels, brain regions and oscillator numbers used in the Cabral model.
\end{flushleft}
\label{tab:brains} 
\end{table}

\bibliographystyle{frontiersinSCNS&ENG} 
\bibliography{bibliothesis}

\end{document}